\documentclass[%
 aip,
 amsmath,amssymb,
 reprint,%
 article-title,
]{revtex4-1}

\usepackage{graphicx}
\usepackage{bm}

\usepackage[utf8]{inputenc}
\usepackage[T1]{fontenc}
\usepackage{mathptmx}
\usepackage{etoolbox}
\usepackage[caption=false]{subfig}
\renewcommand{\bibsection}{\section*{References}}

\makeatletter
\def\@email#1#2{%
 \endgroup
 \patchcmd{\titleblock@produce}
  {\frontmatter@RRAPformat}
  {\frontmatter@RRAPformat{\produce@RRAP{*#1\href{mailto:#2}{#2}}}\frontmatter@RRAPformat}
  {}{}
}%
\makeatother
\begin{document}


\title{On-the-fly $\Delta$-machine learned force fields for warm dense matter using a Yukawa reference model: Application to magnesium}

\author{Sayan Bhowmik}
 \affiliation{College of Engineering, Georgia Institute of Technology, Atlanta, GA 30332, USA}

\author{John E. Pask}
\affiliation{Physics Division, Lawrence Livermore National Laboratory, Livermore, California 94550, USA}%

\author{Phanish Suryanarayana}%
 \email{phanish.suryanarayana@ce.gatech.edu}
  \affiliation{College of Engineering, Georgia Institute of Technology, Atlanta, GA 30332, USA}
   \affiliation{College of Computing, Georgia Institute of Technology, Atlanta, GA 30332, USA}

\date{\today}

\begin{abstract}
We develop a framework for on-the-fly $\Delta$-machine learned force field ($\Delta$-MLFF) molecular dynamics (MD) simulations of warm dense matter (WDM). In particular, we employ a Yukawa reference model that captures the dominant screened-Coulomb interactions between the ions, with its parameters fit on the fly to Kohn--Sham density functional theory (DFT) data, and an MLFF scheme based on the kernel method and Bayesian linear regression that learns the differences in the energies, atomic forces, and stresses between the Yukawa model and Kohn--Sham DFT. We verify the accuracy and performance of the formalism by comparing the predicted properties of warm dense carbon with Kohn--Sham DFT results in the literature. In so doing, we demonstrate that $\Delta$-MLFF has significantly lower errors in the energies, atomic forces, and stresses than the standard MLFF scheme at a comparable cost, while accelerating ab initio MD simulations by up to three orders of magnitude. We apply this framework to calculate the equation of state, structure, and transport properties of magnesium at densities of 5.22 to 10.44 g/cm$^3$ and temperatures of 600 kK to 1.5 MK. We find that the pressure, self-diffusion coefficient, and shear viscosity obey power laws in temperature and density, that the dynamic structure factor exhibits a well-defined ion-acoustic mode whose negative dispersion is characteristic of strongly coupled liquids, and that the Stokes--Einstein relation holds to within 10\% over the conditions considered.
\end{abstract}

\maketitle

\section{\label{sec:intro}Introduction}

Warm dense matter (WDM) refers to the regime that is intermediate between condensed matter and classical plasmas. It is encountered in the interiors of giant planets and stars, along the compression path of inertial confinement fusion (ICF) capsules, and in high energy density (HED) experiments.\cite{graziani2014frontiers} Reliable predictions of its properties are therefore needed, both for the design and interpretation of HED experiments and for modeling planetary and stellar interiors. However, WDM poses significant theoretical challenges, since neither classical plasma physics nor zero-temperature condensed matter theory is applicable: the ions are strongly coupled, the electrons are partially degenerate, and the balance between thermal and quantum effects shifts with density and temperature.

Reliable predictions of WDM properties require a first-principles treatment, for which Kohn--Sham density functional theory (DFT) \cite{kohn1965self, hohenberg1964inhomogeneous} is an attractive choice, with the properties of interest obtained from ab initio molecular dynamics (AIMD) simulations. The cost of Kohn--Sham calculations, however, grows cubically with the number of occupied and partially occupied states, and since this number increases rapidly with temperature, the cubic scaling bottleneck is reached at much smaller system sizes than at ambient conditions. This cost is compounded in AIMD, where the Kohn--Sham equations are solved at every time step, the number of which is dictated by the phenomena of interest and can reach hundreds of thousands, making such simulations prohibitive for all but the smallest systems. Several formulations of Kohn--Sham DFT have been developed to address this limitation, including Spectral Quadrature (SQ) DFT \cite{bhattacharya2022accurate, suryanarayana2013spectral, suryanarayana2018sqdft, pratapa2016spectral}, stochastic DFT (SDFT) \cite{cytter2018stochastic, baer2013self}, mixed stochastic-deterministic DFT (MDFT) \cite{white2020fast}, the density kernel based SQ method (SQ3) \cite{xu2022real}, and spectral partitioned DFT (spDFT) \cite{sadigh2023spectral}. Of these, the SQ method scales linearly with system size and parallelizes efficiently. Moreover, for a given system size, its cost is independent of the number of occupied and partially occupied states, and in fact decreases with increasing temperature. This makes it a particularly attractive choice for WDM studies \cite{bethkenhagen2023properties, zhang2019equation, wu2021development, suryanarayana2025accuracy}.

The cost of AIMD simulations nevertheless remains substantial, which has motivated approximations to Kohn--Sham DFT such as orbital-free molecular dynamics (OFMD) \cite{lambert2006very} and extended first principles molecular dynamics (ext-FPMD) \cite{zhang2016extended, blanchet2021extended}. A different strategy altogether is provided by machine learned force fields (MLFFs) \cite{unke2021machine, poltavsky2021machine, wu2023applications}, which replace the electronic structure calculation at each MD step with an inexpensive model trained on ab initio data, and are being applied to WDM with increasing frequency \cite{hinz2023development, liu2020structure, mahmoud2022predicting, kumar2023transferable, chen2023combining, tanaka2022development, nguyen2024extreme, willman2022machine, berrens2025machine}. However, conventional training requires tens to hundreds of thousands of reference configurations \cite{zhang2020warm, zeng2021ab, liu2021thermal, mahmoud2022predicting}, which are costly to generate and must be regenerated for each new set of conditions. This limitation is overcome by on-the-fly training, in which the model is built and refined during the MD simulation itself, with ab initio calculations performed only when the uncertainty in the model's predictions is too large \cite{jinnouchi2019fly, jinnouchi2020fly, verdi2021thermal, liu2021alpha, Chen2023transferable, kumar2023delta, timmerman2024overcoming}. We have previously combined on-the-fly training, based on the kernel method and Bayesian linear regression, with the SQ method for generating the Kohn--Sham data \cite{kumar2024transport, kumar2024shock}. In so doing, we have shown that AIMD simulations of WDM can be accelerated by up to three orders of magnitude while retaining ab initio accuracy in properties such as the shock Hugoniot, diffusion coefficients, and viscosity.

While the properties obtained from such on-the-fly MLFF simulations are generally in good agreement with AIMD, the accuracy of the underlying energies, atomic forces, and stresses is inherently limited, since the machine-learned model must learn the entire interatomic interaction from a small number of configurations. This is particularly challenging for the strong short-range repulsion between the ions that dominates at WDM conditions, which varies rapidly with the interionic distance and is therefore only sparsely sampled by the training data. Since the resulting errors in the forces and stresses do not necessarily manifest themselves in the averaged properties, they can potentially affect the results in unexpected ways and compromise their accuracy. A natural remedy is $\Delta$-machine learning \cite{ramakrishnan2015big}, in which the machine-learned model learns only the difference between the quantity of interest and a computationally inexpensive baseline that captures the dominant physics. This strategy has been used to obtain Kohn--Sham accuracy from orbital-free DFT \cite{kumar2023delta}, and more generally to improve the data efficiency and stability of machine-learned interatomic potentials \cite{huang2025active, shuaibi2021enabling}. For WDM, the Yukawa model, which describes the screened-Coulomb interaction between the ions with only a few parameters and is widely used as an effective model of ionic interactions in dense plasmas \cite{ohta2000, donko2008, stanton2016}, is a particularly well suited baseline. However, its use within on-the-fly MLFF MD simulations of WDM has not been explored heretofore.

In this work, we develop a framework for on-the-fly $\Delta$-MLFF MD simulations of WDM, wherein a Yukawa model with parameters fit on the fly to Kohn--Sham DFT data serves as the baseline, and an MLFF scheme based on the kernel method and Bayesian linear regression learns the differences in the energies, atomic forces, and stresses between the baseline and Kohn--Sham DFT. Through comparisons with Kohn--Sham results in the literature for warm dense carbon, we show that $\Delta$-MLFF has significantly lower errors in the energies, atomic forces, and stresses than the standard MLFF scheme at a comparable cost, while accelerating AIMD simulations by up to three orders of magnitude. We then apply the framework to calculate the equation of state, structure, and transport properties of magnesium at densities of 5.22 to 10.44 g/cm$^3$ and temperatures of 600 kK to 1.5 MK.

The remainder of this paper is organized as follows. In Section~\ref{sec:formulation}, we describe the formulation and implementation of the $\Delta$-MLFF framework. In Section~\ref{sec:results}, we first verify its accuracy and performance for warm dense carbon, and then apply it to the study of warm dense magnesium. Finally, we provide concluding remarks in Section~\ref{sec:conclusion}.


\section{\label{sec:formulation}Formulation \& Implementation}

\begin{figure*}[htbp]
  \centering
  \includegraphics[width=0.75\textwidth]{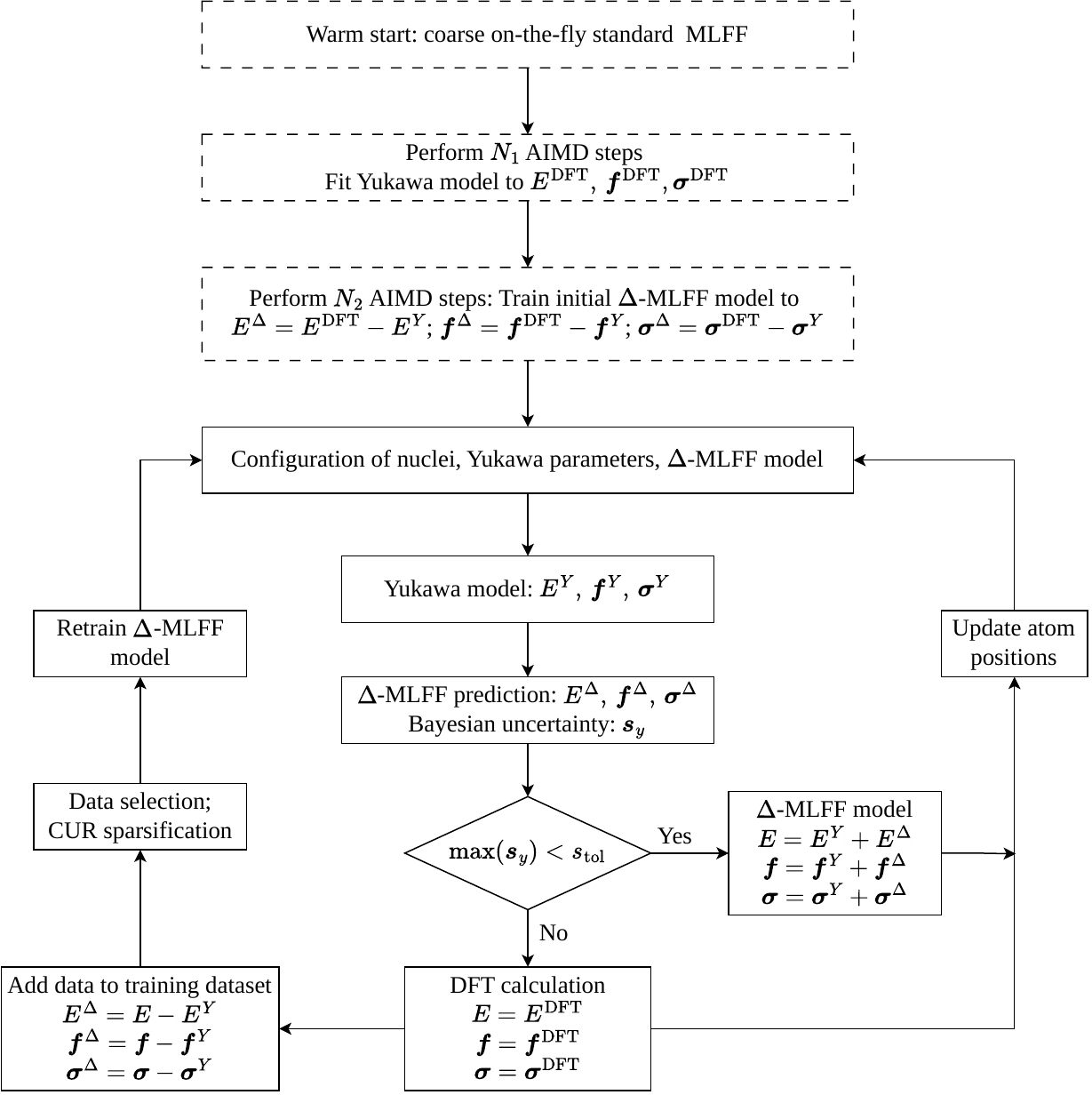}
  \caption{Overview of the on-the-fly $\Delta$-MLFF framework employing the Yukawa model as a baseline for on-the-fly MLFF MD simulations of WDM.}
  \label{fig:flowchart}
\end{figure*}

We have developed the $\Delta$-MLFF framework illustrated in Fig.~\ref{fig:flowchart} for on-the-fly MLFF MD simulations of WDM. In this framework, Kohn--Sham DFT provides the reference values (Sec.~\ref{subsec:DFT}), while a parameter-fitted Yukawa model provides the corresponding baseline values (Sec.~\ref{subsec:yukawa}). The MLFF scheme, based on kernel methods and Bayesian linear regression, learns the difference between the Kohn--Sham and Yukawa values (Sec.~\ref{subsec:delta-MLFF}), and is trained and updated on the fly during the MD simulation (Sec.~\ref{subsec:workflow}). The properties of interest are then calculated from the resulting MD trajectories (Sec.~\ref{subsec:properties}).

The $\Delta$-MLFF framework has been implemented in the real-space electronic structure code SPARC \cite{xu2021sparc, zhang2023version}, extending its on-the-fly MLFF capability \cite{kumar2023delta, kumar2024transport}. The quantities used for both training and prediction are the energy ($E$), atomic forces ($\bm{f}$), and stress tensor ($\bm{\sigma}$), which are decomposed as:
\begin{subequations}\label{eq:split}
\begin{align}
    E           &= E^{\rm Y} + E^{\Delta} \,, \label{eq:e_split} \\
    \bm{f}      &= \bm{f}^{\rm Y} + \bm{f}^{\Delta} \,, \label{eq:f_split} \\
    \bm{\sigma} &= \bm{\sigma}^{\rm Y} + \bm{\sigma}^{\Delta} \,, \label{eq:s_split}
\end{align}
\end{subequations}
where the superscripts ${\rm Y}$ and $\Delta$ denote the Yukawa model and the correction to it, respectively. Since the baseline  accounts for the dominant and rapidly varying part of the interactions between the ions, the residual to be learned is smoother and smaller in magnitude than the original quantities, which can reduce the amount of Kohn--Sham training data required and/or  improve the accuracy of the resulting model. Other choices of baseline are possible for WDM, most notably orbital-free DFT \cite{kumar2023delta}, a simplified form of DFT whose accuracy is known to improve with increasing temperature. However, its use is hampered by practical issues, including the lack of accurate kinetic energy functionals and compatible pseudopotentials. In contrast, the Yukawa baseline is orders of magnitude cheaper to evaluate, while still providing reasonable accuracy in the WDM regime, which motivates its choice in the present work. In what follows, atomic units are used throughout, unless specified otherwise.


\subsection{Kohn--Sham DFT \label{subsec:DFT}}

The finite-temperature Kohn--Sham DFT \cite{Mermin1965} implementation in SPARC is used to generate the training data. In this formalism, the electronic structure is determined by solving the Kohn--Sham eigenvalue problem:
\begin{equation}
    H\psi_n = \lambda_n \psi_n\,,
    \label{eq:KS}
\end{equation}
where $H$ is the Hamiltonian, and $\psi_n$ and $\lambda_n$ are the orbitals and eigenvalues, respectively. Since the Hamiltonian depends on the electron density, which in turn is determined by the Kohn--Sham orbitals, the eigenvalue problem must be solved self-consistently, typically using the self-consistent field (SCF) method.

The conventional approach to solving the Kohn--Sham problem involves diagonalizing the Hamiltonian in each SCF iteration to explicitly compute the occupied and partially occupied Kohn--Sham states. The associated computational cost scales cubically with the number of these states, which increases with electronic temperature, motivating the development of methods that avoid explicit diagonalization. One such approach reformulates the Kohn--Sham problem directly in terms of the finite-temperature density matrix:
\begin{equation}
    D = f(H) = \left(1 + \exp\left(\frac{H - \lambda_F I}{k_B T_e}\right)\right)^{-1}\,,
    \label{eq:DensityMatrix}
\end{equation}
where $f$ is the Fermi--Dirac operator, $I$ is the identity matrix, $k_B$ is the Boltzmann constant, $T_e$ is the electronic temperature, and $\lambda_F$ is the chemical potential, determined by the constraint on the total number of electrons in the simulation cell. In this formulation, the density matrix is computed in each SCF iteration, from which the electron density and related quantities are obtained without explicit calculation of the Kohn--Sham states. This forms the basis of the Gauss variant of the SQ method \cite{suryanarayana2013spectral, suryanarayana2013coarse}, which exploits the nearsightedness of electronic interactions \cite{prodan2005nearsightedness, suryanarayana2017nearsightedness}, manifested in the spatial decay of the density matrix, to express the quantities of interest as localized bilinear forms, or sums thereof, and approximate them using Gauss quadrature. This approach achieves linear scaling with system size and, for a given system size, a computational cost independent of the number of Kohn--Sham states required. The SQ results systematically converge to those obtained by diagonalization as the quadrature order and localization radius are increased, allowing the desired accuracy to be achieved in practice. Moreover, the efficiency of the method improves with increasing electronic temperature, as the density matrix becomes more localized and the Fermi--Dirac operator becomes smoother, making it particularly well suited to WDM conditions, as demonstrated in a number of studies \cite{bethkenhagen2023properties, zhang2019equation, wu2021development, suryanarayana2025accuracy}.

Once self-consistency is achieved, the electronic free energy ($E^{\rm DFT}$), Hellmann--Feynman atomic forces ($\bm{f}^{\rm DFT}$), and Hellmann--Feynman stress tensor ($\bm{\sigma}^{\rm DFT}$) can be evaluated. The diagonalization and SQ approaches are both available in SPARC, allowing seamless interchange between them within the developed $\Delta$-MLFF framework. Only an overview of the Kohn--Sham formalism has been presented here, the expressions for the electron density, energy, atomic forces, and stress tensor, along with the details of their implementation within the real-space finite-difference framework for both approaches, can be found in the literature \cite{ghosh2017sparc2, Sharma2018Stress, suryanarayana2018sqdft, sharma2020real, bhattacharya2022accurate}.


\subsection{Yukawa model\label{subsec:yukawa}}

The Yukawa model is adopted as the baseline, with the interaction between two ions separated by a distance $r$ described by the screened-Coulomb potential:
\begin{equation}
    V^{\rm Y}(r) = c + A\,\frac{e^{-\mu r}}{r} \,,
    \label{eq:yukawa_pair}
\end{equation}
where $A$ is the amplitude, $\mu$ is the inverse screening length, and $c$ is a constant energy shift that accounts for the offset between the Yukawa and DFT energies. This potential arises from the linear response of the electrons screening the Coulomb repulsion between the ions, and is commonly used to study WDM \cite{ohta2000, donko2008, stanton2016}. The Yukawa model is particularly well suited as a baseline for several reasons. First, it captures the dominant repulsive component of the interionic interaction, which varies rapidly with the interionic distance and is particularly challenging for an MLFF to learn from limited training data. Second, the interaction is short-ranged, and its analytic form allows efficient evaluation of the energy, atomic forces, and stresses. Third, the model involves only three parameters, $A$, $\mu$, and $c$, which can be reliably determined from a small number of Kohn--Sham calculations, as described below. The corresponding energy ($E^{\rm Y}$), atomic forces ($\bm{f}^{\rm Y}$), and stress tensor ($\bm{\sigma}^{\rm Y}$) can then be readily evaluated.

The parameters $A$, $\mu$, and $c$ are determined by fitting the Yukawa model to the energy, atomic forces, and stresses from DFT, i.e., $E^{\rm DFT}$, $\bm{f}^{\rm DFT}$, and $\bm{\sigma}^{\rm DFT}$. In particular, we minimize the objective function:
\begin{equation}
    J(\bm{\theta}, \mu) = \left\| \bm{W} \left( \bm{y}^{\rm Y}(\bm{\theta}, \mu) - \tilde{\bm{y}}^{\rm DFT} \right) \right\|^2 \,,
    \label{eq:obj}
\end{equation}
where $\bm{\theta} = (A, c)^{\rm T}$ collects the parameters that enter linearly; $\tilde{\bm{y}}^{\rm DFT}$ is the vector containing $E^{\rm DFT}$, $\bm{f}^{\rm DFT}$, and $\bm{\sigma}^{\rm DFT}$ for the configurations used in the fitting; $\bm{y}^{\rm Y}$ is the corresponding vector containing $E^{\rm Y}$, $\bm{f}^{\rm Y}$, and $\bm{\sigma}^{\rm Y}$ from the Yukawa model; and $\bm{W}$ is a diagonal matrix of weights, with entries $1/\sqrt{N}$ for the energies, $1/\sqrt{3N}$ for the components of the forces, and $1$ for the components of the stresses, where $N$ is the number of atoms. The parameters are determined using the variable projection method \cite{golub1973}.

The framework extends naturally to multicomponent systems. In this case, the interaction between each pair of species is described by a Yukawa potential of the form in Eq.~\eqref{eq:yukawa_pair}, with the amplitude specific to the species pair, while the inverse screening length and the energy shift remain common to all pairs. The resulting parameters are again determined by minimizing the objective function in Eq.~\eqref{eq:obj}, with $\bm{\theta}$ now collecting the amplitudes for all species pairs together with the energy shift. The variable projection method again applies, since the amplitudes and energy shift enter linearly, with the inverse screening length remaining the only nonlinear parameter. Further generalizations, such as a pair-specific inverse screening length, are possible; however, since the residual is learned by the MLFF, such refinements of the baseline are not expected to be necessary in practice.


\subsection{$\Delta$-MLFF model \label{subsec:delta-MLFF}}

The machine-learned model for the correction to the Yukawa energy in Eq.~\eqref{eq:e_split} takes the form \cite{jinnouchi2020fly, kumar2023delta}:
\begin{subequations}
\begin{align}
    E^{\Delta} &= \sum_{e}\sum_{i=1}^{N_A^e}\varepsilon_i^e = \sum_{e}\sum_{i=1}^{N_A^e}\sum_{t=1}^{N_T^e}\tilde{w}_t^e\, k(\bm{x}_i^e,\tilde{\bm{x}}_t^e) \,, \label{eq:kernel}\\
    k(\bm{x}_i^e,\tilde{\bm{x}}_t^e) &= \left(\frac{\bm{x}_i^e\cdot\tilde{\bm{x}}_t^e}{|\bm{x}_i^e||\tilde{\bm{x}}_t^e|}\right)^{\xi} \,, \label{eq:polykernel}
\end{align}
\end{subequations}
where $e$ denotes the chemical element that also serves as an index, $\varepsilon_i^e$ are the atomic energies, $\tilde{w}_t^e$ are the model weights, and $k(\bm{x}_i^e,\tilde{\bm{x}}_t^e)$ is the polynomial kernel of degree $\xi$ that measures the similarity of the Smooth Overlap of Atomic Positions (SOAP) descriptors \cite{bartok2013representing} $\bm{x}_i^e$ and $\tilde{\bm{x}}_t^e$, the former corresponding to the $N_A^e$ atoms in the current atomic configuration, and the latter corresponding to the $N_T^e$ atoms in the training dataset. Since the dependence of the descriptors on the atomic positions is explicitly known, the corresponding atomic forces $\bm{f}^{\Delta}$ and stress tensor $\bm{\sigma}^{\Delta}$ are immediately accessible, detailed expressions for which can be found in the literature \cite{kumar2023delta}. The model is trained such that its predictions $E^{\Delta}$, $\bm{f}^{\Delta}$, and $\bm{\sigma}^{\Delta}$ match the differences between the DFT and Yukawa energies, atomic forces, and stresses, respectively. 

The weights are determined through Bayesian linear regression:
\begin{equation}
    \tilde{\bm{w}} = \beta\left(\alpha\bm{I} + \beta\tilde{\bm{K}}^{\rm T}\tilde{\bm{K}}\right)^{-1}\tilde{\bm{K}}^{\rm T}\tilde{\bm{y}} \,,
    \label{eq:blr}
\end{equation}
where $\tilde{\bm{w}}$ is the vector of weights; $\alpha$ and $\beta$ are parameters that are determined by maximizing the evidence function; $\tilde{\bm{y}}$ is the vector containing the differences between the DFT and Yukawa energies, atomic forces, and stresses --- suitably normalized by their mean and standard deviation values \cite{kumar2023delta} --- for the atomic configurations in the training dataset; and $\tilde{\bm{K}}$ is the corresponding covariance matrix. The machine-learned correction for a given atomic configuration, collected in the vector $\bm{y}$, can then be predicted using the relation $\bm{y} = \bm{K}\tilde{\bm{w}}$, where $\bm{K}$ is the covariance matrix associated with the given atomic configuration. The uncertainty in the values so predicted is ascertained using the relation:
\begin{equation}
    \bm{s}_{\bm{y}}^2 = {\rm diag}\left(\frac{1}{\beta}\bm{I} + \bm{K}\left(\alpha\bm{I} + \beta\tilde{\bm{K}}^{\rm T}\tilde{\bm{K}}\right)^{-1}\bm{K}^{\rm T}\right) \,,
    \label{eq:uncertainty}
\end{equation}
where $\bm{s}_{\bm{y}}$ is a vector with the Bayesian error in the energy, atomic forces, and stresses as entries. Since the Yukawa model is deterministic, the Bayesian errors in $E$, $\bm{f}$, and $\bm{\sigma}$ are identical to those in $E^{\Delta}$, $\bm{f}^{\Delta}$, and $\bm{\sigma}^{\Delta}$, respectively.


\subsection{On-the-fly MLFF MD \label{subsec:workflow}}

The MD simulation begins with a warm-start stage, in which a coarse on-the-fly MLFF simulation generates a physically reasonable and decorrelated set of initial configurations. A relatively large threshold $s_{\rm tol}$ is employed at this stage to minimize the number of Kohn--Sham calculations. The coarse MLFF model is then discarded, and $N_1$ AIMD steps are performed to generate the data used to fit the Yukawa parameters. With these parameters fixed, a further $N_2$ AIMD steps are performed to generate the initial training dataset, which consists of the differences between the DFT and Yukawa quantities. The initial $\Delta$-MLFF model is then constructed from this dataset.

The subsequent MD steps are performed using the $\Delta$-MLFF framework. At each step, the Yukawa model provides $E^{\rm Y}$, $\bm{f}^{\rm Y}$, and $\bm{\sigma}^{\rm Y}$, while the $\Delta$-MLFF model provides $E^{\Delta}$, $\bm{f}^{\Delta}$, and $\bm{\sigma}^{\Delta}$, along with the Bayesian error $\bm{s}_{\bm{y}}$. The total energy $E$, atomic forces $\bm{f}$, and stress tensor $\bm{\sigma}$ are then assembled through Eq.~\eqref{eq:split}, provided that the maximum Bayesian error in the forces remains below the threshold, i.e., $\max(\bm{s}_{\bm{y}}) < s_{\rm tol}$. Otherwise, a Kohn--Sham calculation is performed, and $E = E^{\rm DFT}$, $\bm{f} = \bm{f}^{\rm DFT}$, and $\bm{\sigma} = \bm{\sigma}^{\rm DFT}$ are used directly. In this case, the differences between the DFT and Yukawa quantities are added to the training dataset, and the $\Delta$-MLFF model is retrained. In either case, the atomic positions are then updated using the resulting forces.

The threshold $s_{\rm tol}$ is not fixed, but is instead set to the maximum Bayesian error in the forces at the MD step following each training step, scaled by a factor slightly larger than unity. To avoid the cubic scaling bottleneck in training, a two-step data selection procedure is adopted: only those atoms whose Bayesian error in the forces exceeds $s_{\rm tol}$ are added to the training dataset, and CUR \cite{mahoney2009cur} is then applied to the resulting dataset for downsampling.

\subsection{\label{subsec:properties}Properties}

The properties of interest are calculated from the MD trajectories. In this work, these are the pressure, which provides the equation of state (EOS); the Coulomb coupling and electron degeneracy parameters, which characterize the plasma state; the static and dynamic ion structure factors, which characterize the ionic structure and its collective dynamics; and the self-diffusion coefficient and shear viscosity, which characterize the ionic transport. In what follows, $\langle \cdot \rangle$ denotes the ensemble average.

The Coulomb coupling parameter and the electron degeneracy parameter are calculated as \cite{graziani2014frontiers}:
\begin{equation}
    \Gamma = \left(\frac{4\pi n_{i}}{3}\right)^{1/3} \frac{{Z^*}^2}{k_B T_{i}} \,, \quad \Theta = \frac{2 k_B T_{e}}{(3\pi^2 n_{e})^{2/3}} \,,
    \label{eq:plasma_params}
\end{equation}
where $Z^*$ is the average ionization of the ions, calculated using the density of states (DOS) integration method \cite{callow2023improved}; $n_{i} = N/V$ and $n_{e}$ are the ion and electron number densities, with $N$ the number of atoms and $V$ the cell volume; and $T_{i}$ and $T_{e}$ are the ionic and electronic temperatures.

The static and dynamic ion structure factors are calculated as \cite{rapaport2004art}:
\begin{equation}
    S(\bm{q}) = F(\bm{q}, 0) \,, \quad S(\bm{q}, \omega) = \frac{1}{2\pi}\int_{-\infty}^{\infty}\!dt\, e^{\mathrm{i}\omega t} F(\bm{q}, t) \,,
    \label{eq:S}
\end{equation}
where $F(\bm{q}, t) = \frac{1}{N}\left\langle \rho_{\bm{q}}(0)^{*}\,\rho_{\bm{q}}(t) \right\rangle$ is the intermediate scattering function, $\rho_{\bm{q}}(t) = \sum_{i=1}^{N} e^{-\mathrm{i}\bm{q}\cdot\bm{R}_i(t)}$ is the Fourier component of the ion density, $\bm{R}_i$ are the ionic positions, and $\bm{q}$ are the wavevectors commensurate with the periodic cell. Since the systems considered here are assumed to be isotropic, results are averaged over all $\bm{q}$ with the same magnitude $q = |\bm{q}|$, giving $S(q)$ and $S(q, \omega)$.

The pressure is calculated as:
\begin{equation}
    P = \frac{N k_B T_{i}}{V} - \frac{1}{3}\left\langle {\rm tr}\,\bm{\sigma} \right\rangle \,,
    \label{eq:P}
\end{equation}
where the first term is the ideal gas contribution, and the second is the contribution from the interactions, with the stress tensor $\bm{\sigma}$ given by $\bm{\sigma}^{\rm DFT}$ at Kohn--Sham steps and by Eq.~\eqref{eq:split} at $\Delta$-MLFF steps.

The self-diffusion coefficient and shear viscosity are calculated from the Green--Kubo relations \cite{rapaport2004art, alfe1998first}:
\begin{align}
    D &= \frac{1}{3N}\int_0^{\infty}\!dt\,\sum_{i=1}^{N}\langle\bm{v}_i(0)\cdot\bm{v}_i(t)\rangle \,, \label{eq:D}\\
    \eta &= \frac{V}{5k_B T_{i}}\int_0^{\infty}\!dt\,\sum_{m=1}^{5}\langle\sigma'_m(0)\,\sigma'_m(t)\rangle \,, \label{eq:eta}
\end{align}
where $\bm{v}_i$ are the ionic velocities, and $\sigma'_m$ are the five independent components of the deviatoric stress, $\bm{\sigma}' = [\sigma_{12}, \sigma_{23}, \sigma_{31}, (\sigma_{11}-\sigma_{22})/2, (\sigma_{22}-\sigma_{33})/2]$.

The statistical error in each property is estimated using the blocking method \cite{flyvbjerg1989}. For $D$ and $\eta$, the method is applied to the values obtained from the individual time origins, and the reported value is their mean.

\section{\label{sec:results}Results \& Discussion}
In this section, we first assess the accuracy and performance of the developed $\Delta$-MLFF framework for warm dense carbon (Sec.~\ref{subsec:accuracyPerformance}) and subsequently apply it to warm dense magnesium (Sec.~\ref{subsec:Mg}).

In all simulations, the MD runs are performed in the NVK ensemble using a Gaussian thermostat~\cite{minary2003algorithms}, with the electronic and ionic temperatures set equal: $T = T_e = T_i$. Each run is initialized from a random packing of ions at the chosen density. The time step is chosen to scale as $\Delta t \propto M^{5/6}\rho^{-1/3}T^{-1/2}$, where $M$ is the ionic mass and $\rho$ is the mass density~\cite{zhang2016extended, hou2021ionic}. A total of 500\,000 MD steps are performed, the first 5000 of which are discarded as equilibration. The resulting statistical errors in the pressure $P$, self-diffusion coefficient $D$, and shear viscosity $\eta$ are within 4.4~GPa, $2.5 \times 10^{-4}$~cm$^2$/s, and 1.3~mPa\,s, respectively, corresponding to relative errors of 0.01\%, 0.5\%, and 5\%.

The Kohn--Sham DFT calculations employ the local density approximation (LDA)~\cite{kohn1965self}, which provides sufficient accuracy for the conditions considered~\cite{suryanarayana2025accuracy, berrens2025machine}, and ONCV pseudopotentials~\cite{hamann2013optimized} with suitably small cutoff radii and appropriate numbers of electrons in valence~\cite{suryanarayana2025accuracy, kumar2023delta, kumar2024transport}. The grid spacing, SQ truncation radius, SQ quadrature order, and other numerical parameters are converged at each $\rho$--$T$ state point such that the discretization error in the maximum force component and the pressure is below 1\%.

The warm start consists of a coarse on-the-fly MLFF simulation of 100 MD steps with a scaling factor of 1.005 for $s_{\rm tol}$, followed by $N_1 = 5$ AIMD steps for fitting the Yukawa model and $N_2 = 15$ AIMD steps for training the initial $\Delta$-MLFF model. The Yukawa model employs an interaction cutoff of 15 bohr. The ML hyperparameters are chosen to be the same as in earlier work~\cite{kumar2024transport}: SOAP descriptors with a maximum angular order of 6, 8 radial basis functions, and a cutoff radius of 10 bohr; a scaling factor of 1.001 for $s_{\rm tol}$; and a regularization parameter of $10^{-14}$.

Detailed data for all results presented in this work is provided in the Supplementary Material.

\subsection{\label{subsec:accuracyPerformance}Accuracy \& performance: warm dense carbon}

\begin{figure}[htbp]
    \centering
    \includegraphics[width=0.8\linewidth]{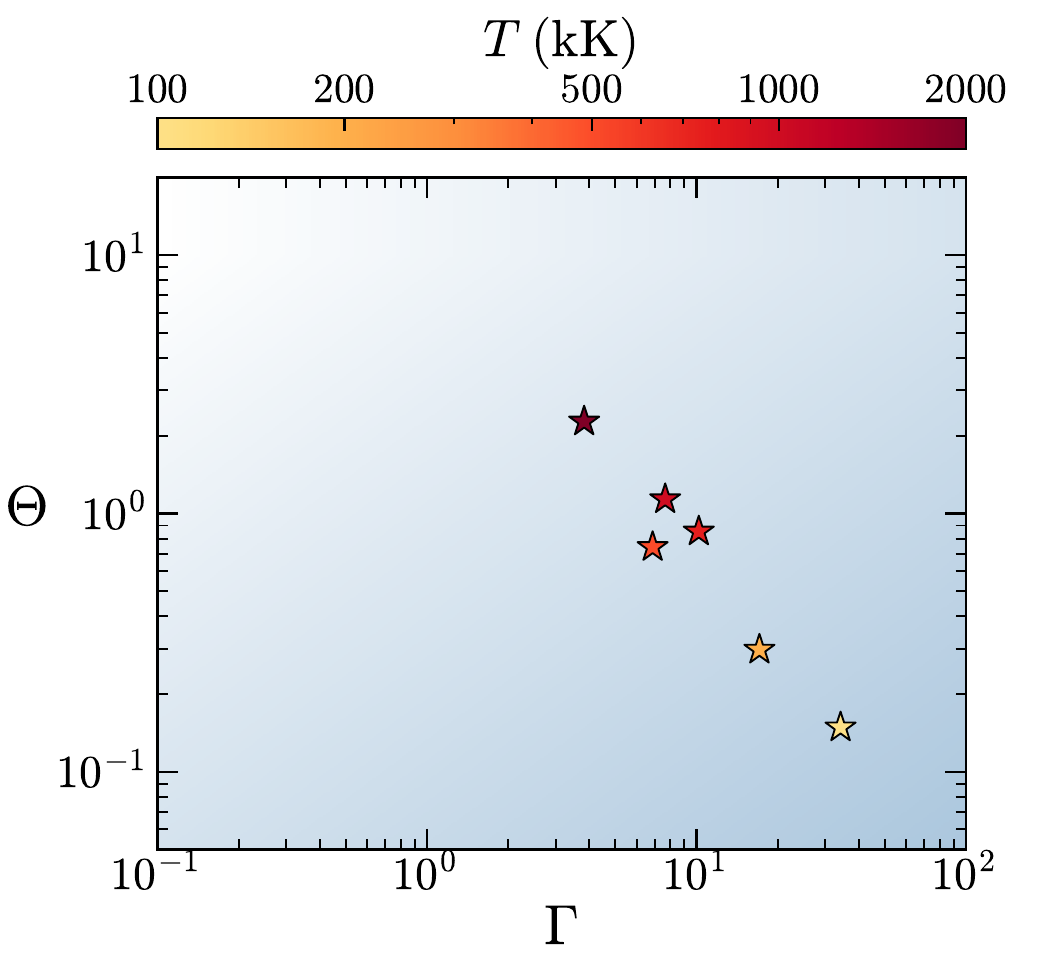}
    \caption{Electron degeneracy parameter ($\Theta$) and ionic coupling parameter ($\Gamma$) for  carbon at the densities and temperatures considered in this work.}
    \label{fig:theta_gamma_C}
\end{figure}

We consider carbon at a density of $\rho = 10$~g/cm$^3$ and temperatures of $T=100$~kK, 200 kK, 500 kK, 750 kK, 1 MK, and 2 MK. These $\rho$--$T$ states are chosen because reference DFT data are available for them~\cite{bethkenhagen2023properties}, allowing for the accuracy of the $\Delta$-MLFF scheme to be assessed. They lie in the WDM regime, with plasma parameters of $\Theta = 0.15$--$2.3$ and $\Gamma = 3.8$--$34$ (Fig.~\ref{fig:theta_gamma_C}), indicative of partially degenerate electrons and strongly coupled ions, thereby motivating the need for ab initio accuracy~\cite{stanek2024review}.

The carbon simulations use cells containing $N = 200$, 169, 127, 109, 95, and 64 atoms at 100 kK, 200 kK, 500 kK, 750 kK, 1 MK, and 2 MK, respectively. The time step decreases from 0.053 fs at 100 kK to 0.012 fs at 2 MK. All six electrons of carbon are treated as valence, using an ONCV pseudopotential with a cutoff radius of 0.63 bohr and two projectors per angular momentum channel. The electronic structure is computed using diagonalization at 100 kK and 200 kK, and using the SQ method at the remaining higher temperatures. The grid spacing is 0.13 bohr, and in the SQ calculations the quadrature order ranges from 65 to 30 and the truncation radius from 1.5 to 1.0 bohr, with the first value of each range corresponding to 500 kK and the second to 2 MK. 

For comparison, we also perform simulations using the standard MLFF scheme, i.e., with $E^{\rm Y} \equiv 0$, employing the same hyperparameters as in $\Delta$-MLFF. To isolate the contribution of the Yukawa model, we additionally perform MD using the standalone Yukawa model, with the parameters $A$, $\mu$, and $c$ taken from the corresponding $\Delta$-MLFF simulation. These simulations are identical to the corresponding $\Delta$-MLFF simulations, with the same number of atoms, cell, time step, thermostat, initial positions, and number of MD steps.

\subsubsection{Accuracy \label{subsec:accuracy}}

We now assess the accuracy of the $\Delta$-MLFF scheme against DFT. In particular, we first compare the energy, atomic forces, and stresses at individual MD snapshots. Second, we compare the radial distribution function, pressure, self-diffusion coefficient, and shear viscosity obtained from the MD runs.
\paragraph{Energy, forces, and stresses}

\begin{figure*}[htbp]
    \centering
    \subfloat[]{\includegraphics[width=0.4\textwidth]{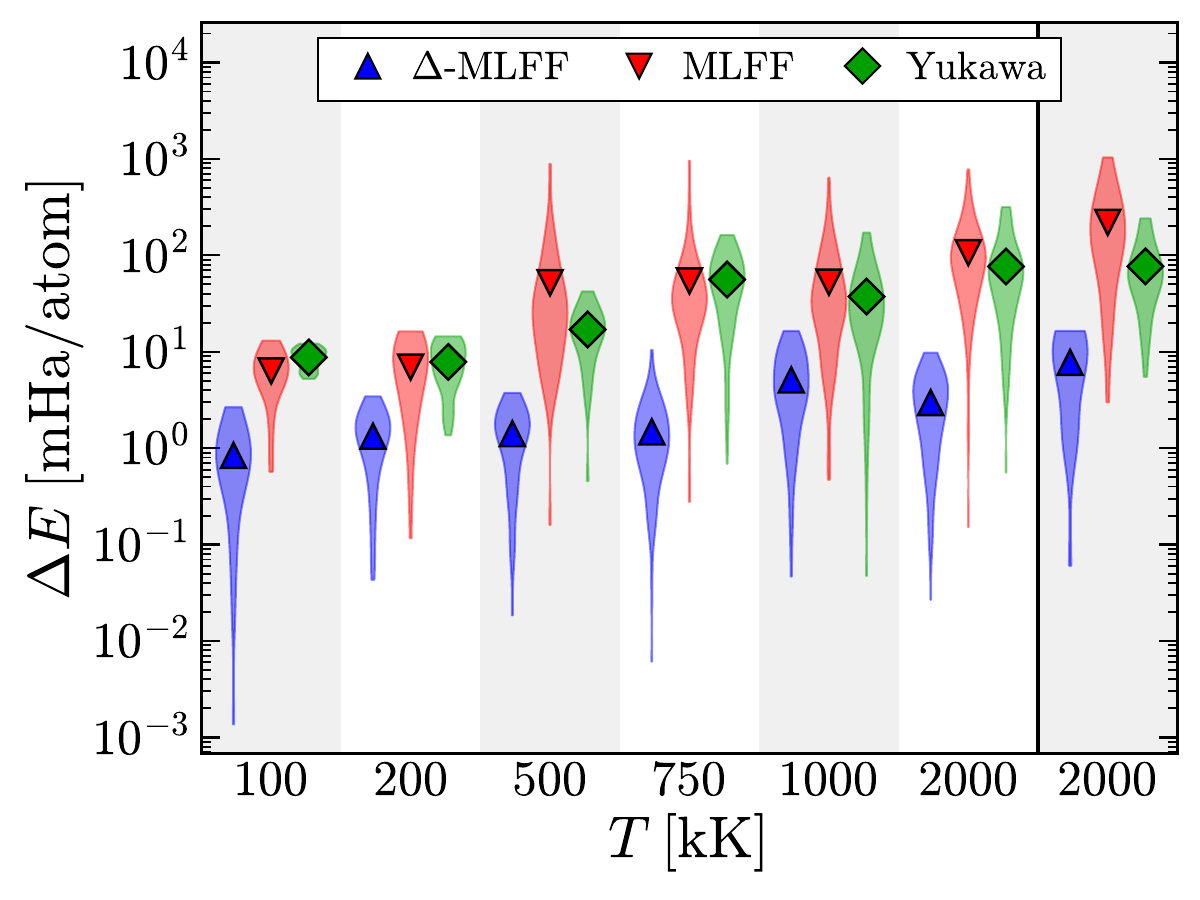}\label{fig:C_err_E}}
    \subfloat[]{\includegraphics[width=0.4\textwidth]{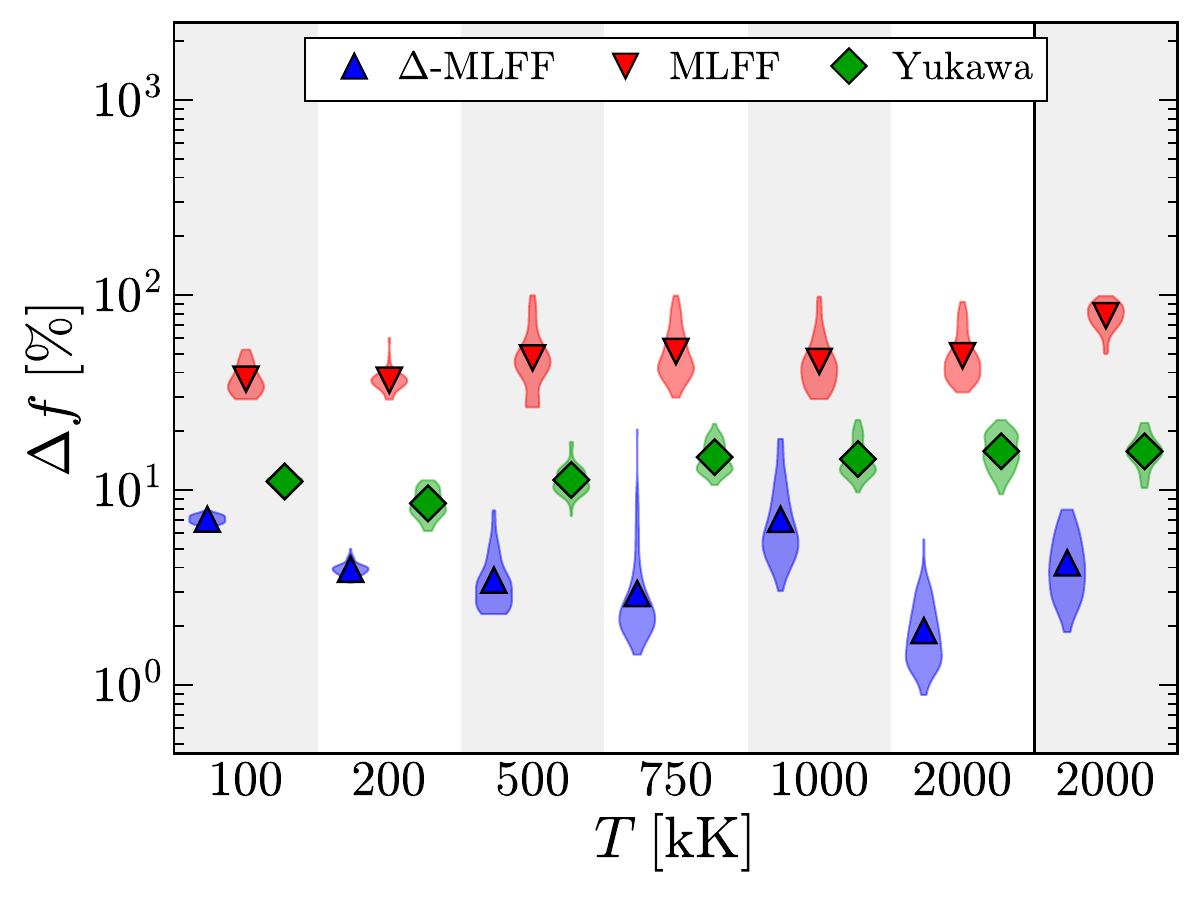}\label{fig:C_err_F}}\\
    \subfloat[]{\includegraphics[width=0.4\textwidth]{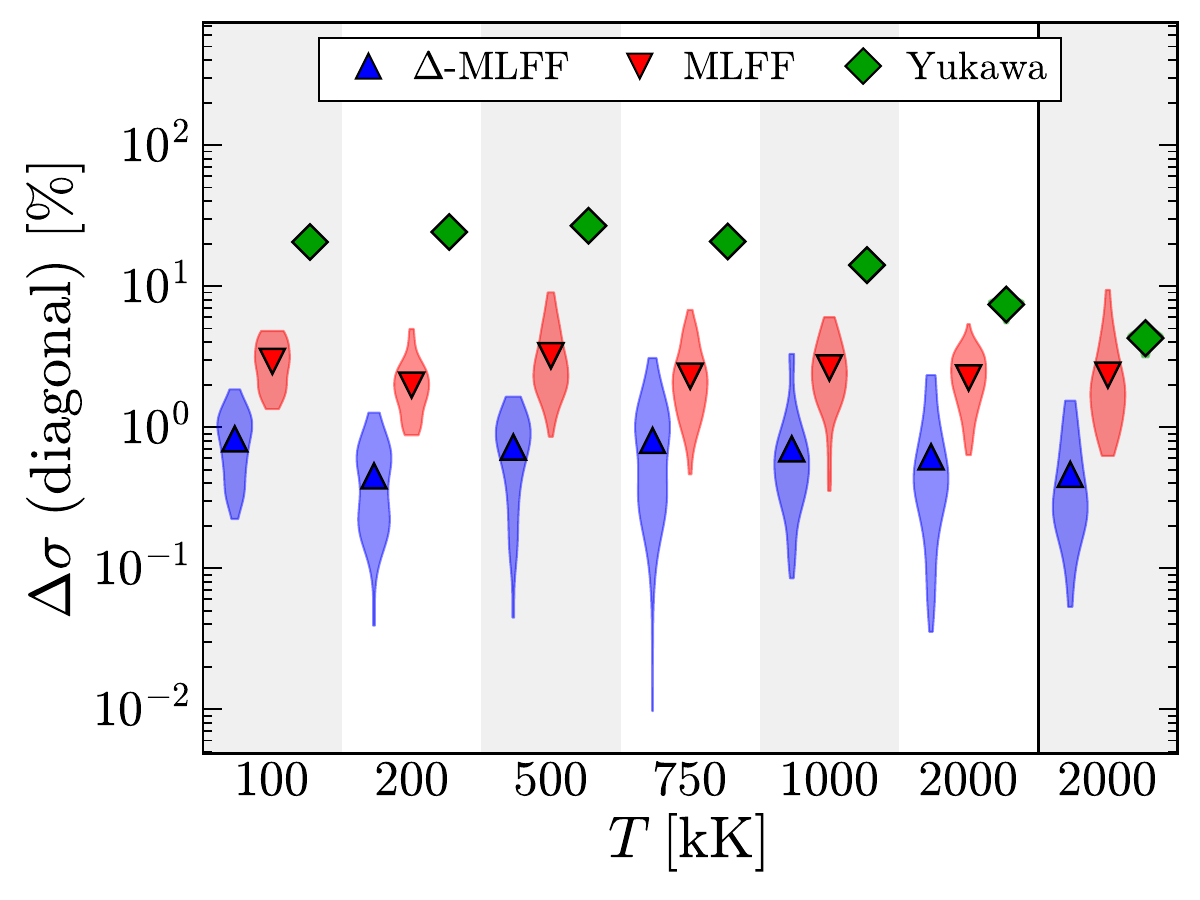}\label{fig:C_err_Sdiag}}
    \subfloat[]{\includegraphics[width=0.4\textwidth]{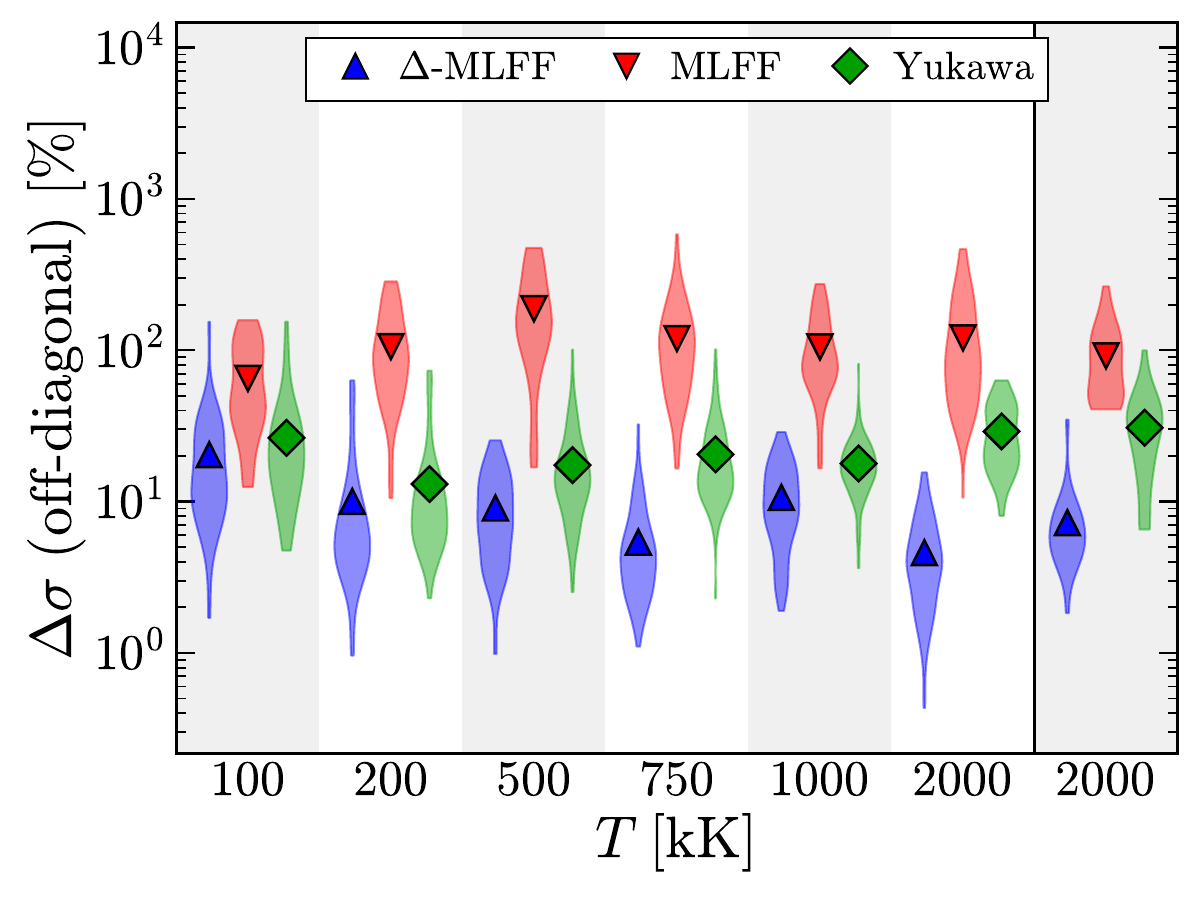}\label{fig:C_err_Soff}}
    \caption{Errors of $\Delta$-MLFF, MLFF, and the Yukawa model with respect to DFT for carbon at 10~g/cm$^3$ and, to the right of the solid vertical line, at 16~g/cm$^3$ and 2 MK. (a) Absolute error in the free energy per atom, $|\Delta E|$. (b) Relative error (\%) in the forces, $\Delta f$, defined as the magnitude of the difference between the predicted and DFT force vectors on all atoms, divided by the magnitude of the DFT force vector. (c), (d) Relative errors (\%) in the diagonal and off-diagonal components of the stress tensor, respectively, $\Delta\sigma$, defined analogously to $\Delta f$.}
    \label{fig:C_err}
\end{figure*}

In Fig.~\ref{fig:C_err}, we present the errors in the energy, forces, and stresses with respect to DFT at the DFT calls of the on-the-fly MLFF MD. The standalone Yukawa model is evaluated on the same configurations as $\Delta$-MLFF, whereas MLFF is evaluated at its own DFT calls.

We observe that $\Delta$-MLFF has the lowest mean error for every quantity at every state. In particular, its mean energy error is 0.8--7.7~mHa/atom, compared with 6.3--220~mHa/atom for MLFF and 7.8--77~mHa/atom for the standalone Yukawa model (Fig.~\ref{fig:C_err}(a)). Notably, the largest errors of MLFF reach 1000~mHa/atom at $T \geq 500$~kK, whereas those of $\Delta$-MLFF remain below 17~mHa/atom. We also observe that the mean force error of $\Delta$-MLFF is 1.9--7.1\%, which is 5--26 times smaller than the 36--78\% of MLFF (Fig.~\ref{fig:C_err}(b)). MLFF performs worst at 16~g/cm$^3$, where its mean energy and force errors of 220~mHa/atom and 78\% are 2.1 and 1.6 times those at 10~g/cm$^3$, respectively, whereas those of $\Delta$-MLFF remain at 7.7~mHa/atom and 4.2\%. It is worth noting that the standalone Yukawa model already achieves mean force errors of 8.5--16\%, whereby the machine-learned correction needs to learn only the difference between the DFT and Yukawa forces, which is about an order of magnitude smaller than the forces themselves. Indeed, the correction lowers the force error of the Yukawa model by a factor of 1.6--8.

For the diagonal stress tensor components, we find that the mean errors are 0.45--0.82\% for $\Delta$-MLFF and 2.0--3.2\% for MLFF (Fig.~\ref{fig:C_err}(c)). The standalone Yukawa model has the largest errors, 4.3--27\%, which do not vary noticeably from one configuration to the next. These errors correspond to the part of the pressure that is independent of the ion positions and is therefore missing from the Yukawa model, which the machine-learned correction reduces to below 1\%. For the off-diagonal stress tensor components, which enters the calculation of the viscosity, we find that the mean errors are 4.6--20\% for $\Delta$-MLFF, 13--31\% for the standalone Yukawa model, and 65--188\% for MLFF (Fig.~\ref{fig:C_err}(d)). In fact, at 10~g/cm$^3$ and $T \geq 200$~kK, the shear-stress errors of MLFF are on average larger than the shear stresses themselves.

In summary, $\Delta$-MLFF has the lowest errors in all quantities at all states. Of the other two methods, the standalone Yukawa model performs better for the forces and the off-diagonal stress tensor components, MLFF performs better for the diagonal stress tensor components, and their energy errors are comparable.

\paragraph{Radial distribution function}

\begin{figure*}[htbp]
    \centering
    \subfloat[]{\includegraphics[width=0.32\textwidth]{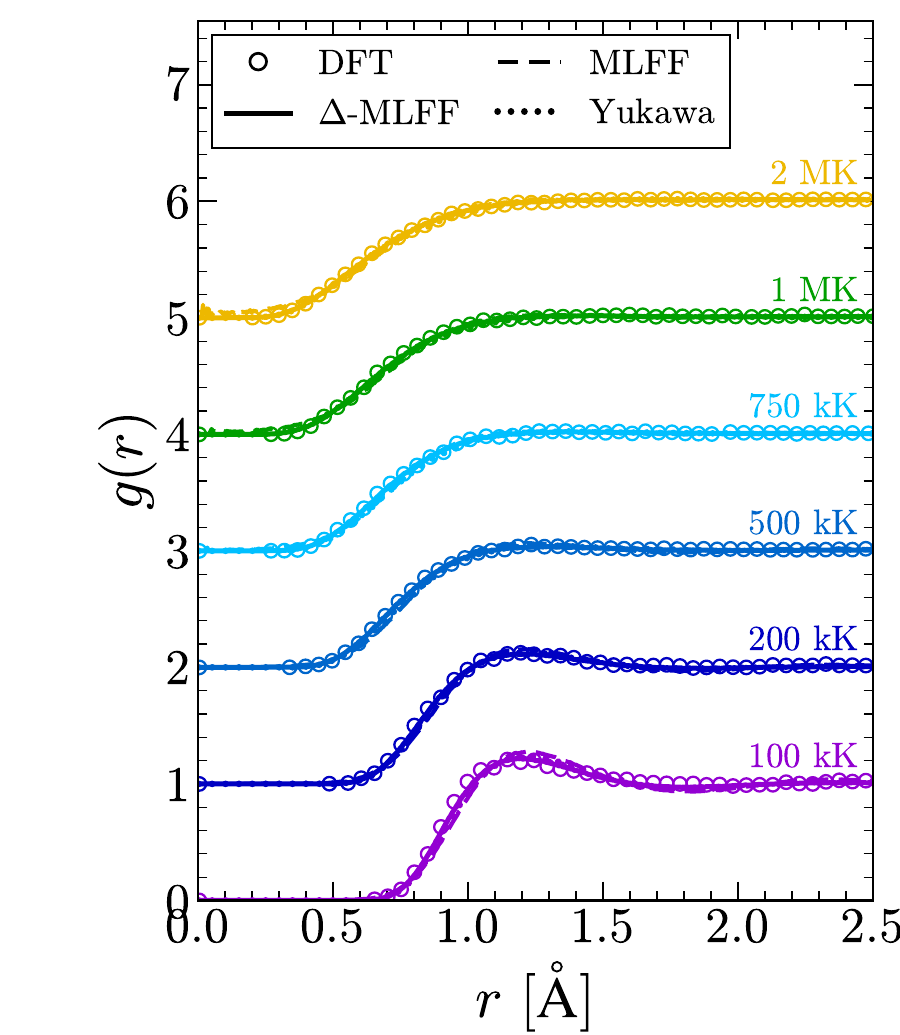}\label{fig:C_gr_offset}}\hfill
    \subfloat[]{\includegraphics[width=0.32\textwidth]{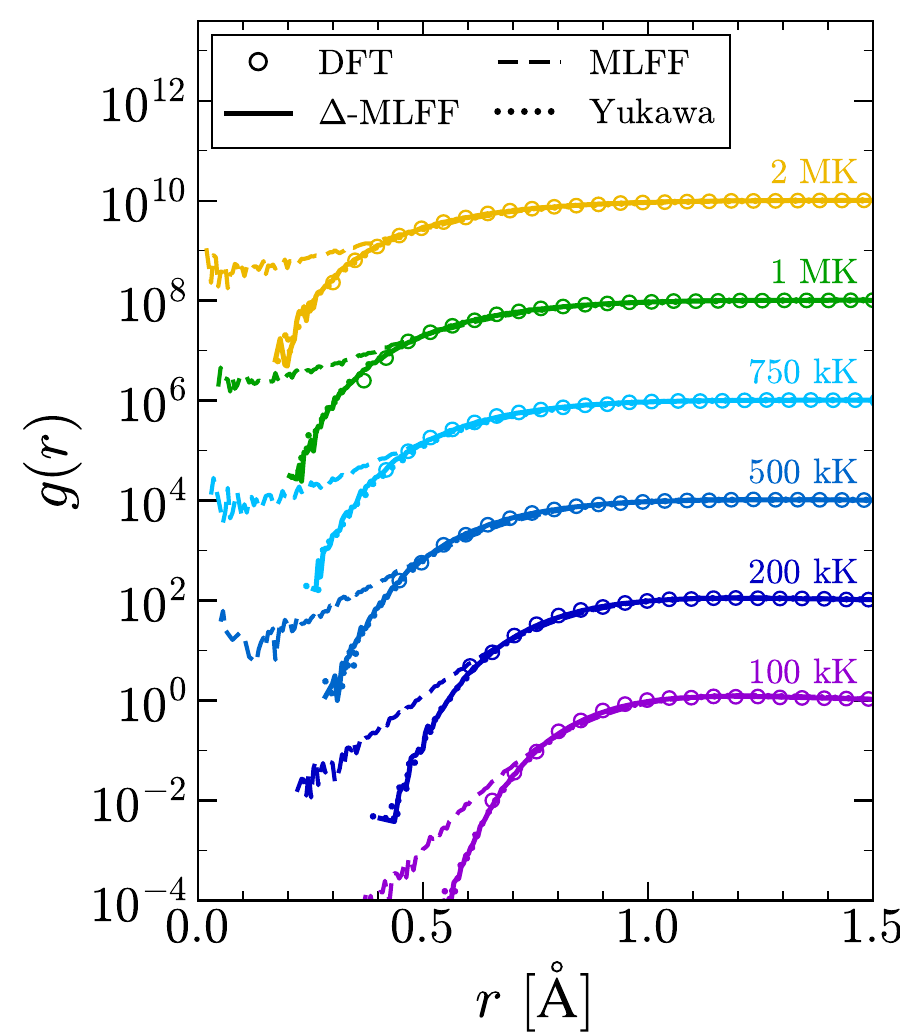}\label{fig:C_gr_log}}\hfill
    \subfloat[]{\includegraphics[width=0.32\textwidth]{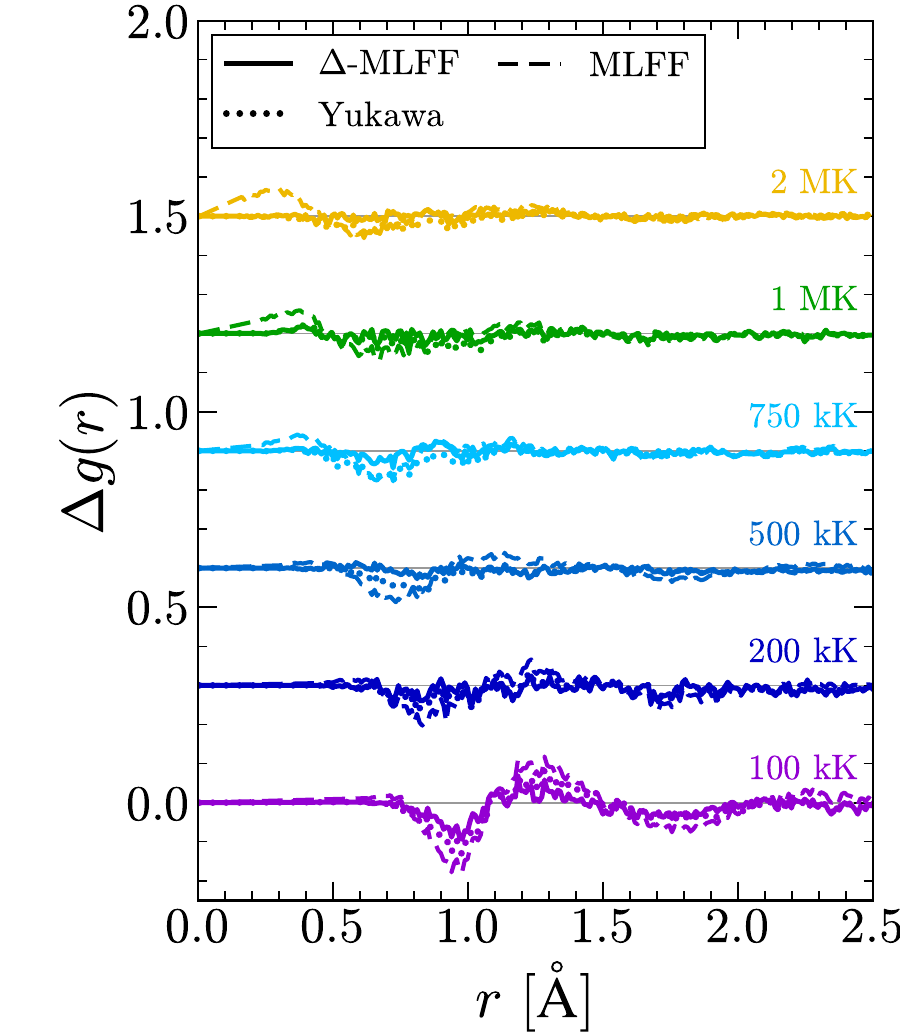}\label{fig:C_gr_diff}}
    \caption{Radial distribution function $g(r)$ for carbon at 10~g/cm$^3$ obtained from $\Delta$-MLFF, MLFF, and the Yukawa model, compared with DFT~\cite{bethkenhagen2023properties}. (a) $g(r)$ on a linear scale. (b) $g(r)$ on a logarithmic scale. (c) Deviation of $g(r)$ from DFT. Results at different temperatures are shifted for clarity.}
    \label{fig:C_gr}
\end{figure*}

In Fig.~\ref{fig:C_gr}, we compare the radial distribution function $g(r)$ of carbon with DFT~\cite{bethkenhagen2023properties}. We observe that $\Delta$-MLFF reproduces the DFT $g(r)$ at all six temperatures, with root-mean-square (RMS) deviations of 0.007--0.027 over $r \leq 2.5$~\AA\ (Fig.~\ref{fig:C_gr}(c)). Notably, this is comparable to the RMS deviations of 0.011--0.017 between the different DFT codes in Ref.~\onlinecite{bethkenhagen2023properties} at 100 kK and 200 kK, which reflect the statistical noise of the 64-atom reference simulations. In particular, the first peak at 100 kK ($g = 1.22$ at $r = 1.18$~\AA) matches that of DFT ($g = 1.21$ at $r = 1.15$~\AA), and the distance at which $g(r)$ reaches 1/2 differs from DFT by only 0.002--0.011~\AA.

In contrast, MLFF has 2--4 times larger RMS deviations of 0.023--0.061. Its $g(r)$ rises 0.027--0.032~\AA\ farther out at all temperatures, and its first peak at 100 kK is too high ($g = 1.28$ at $r = 1.23$~\AA). We also observe that MLFF allows the ions to approach each other unphysically closely (Fig.~\ref{fig:C_gr}(b)). Indeed, at $T \geq 500$~kK, its $g(r)$ exceeds $0.001$ already at $r = 0.02$--0.05~\AA, compared with 0.18--0.34~\AA\ for $\Delta$-MLFF, which follows the DFT data down to $g \approx 0.01$. Since the training data contain no configurations with such small interatomic distances, the machine-learned model extrapolates there without any short-range repulsion. In $\Delta$-MLFF, this repulsion is provided by the Yukawa model.

The standalone Yukawa model yields RMS deviations of 0.014--0.040, intermediate between those of $\Delta$-MLFF and MLFF. Its $g(r)$ departs from zero at the same distance as that of $\Delta$-MLFF, confirming that the Yukawa model sets the short-range repulsion. However, its rise is shifted outward by 0.014--0.027~\AA, and its first peak at 100 kK is too high ($g = 1.24$ at $r = 1.23$~\AA). We therefore conclude that the machine-learned correction mainly improves the structure around the first peak, where a screened Coulomb interaction cannot capture the effects of the electronic structure. It is worth noting that the remaining differences between $\Delta$-MLFF and DFT are consistent with the differences in cell size (64 atoms in Ref.~\onlinecite{bethkenhagen2023properties} versus 64--200 here, which shifts the large-$r$ limit of $g(r)$ by about $1/N$), trajectory length, and pseudopotential (only the 2s and 2p electrons are treated as valence below 750 kK in Ref.~\onlinecite{bethkenhagen2023properties}).

In summary, $\Delta$-MLFF agrees best with DFT at every temperature (RMS deviation 0.007--0.027), followed by the standalone Yukawa model (0.014--0.040) and then MLFF (0.023--0.061).

\paragraph{Pressure}

\begin{figure}[htbp]
    \centering
    \includegraphics[width=0.8\linewidth]{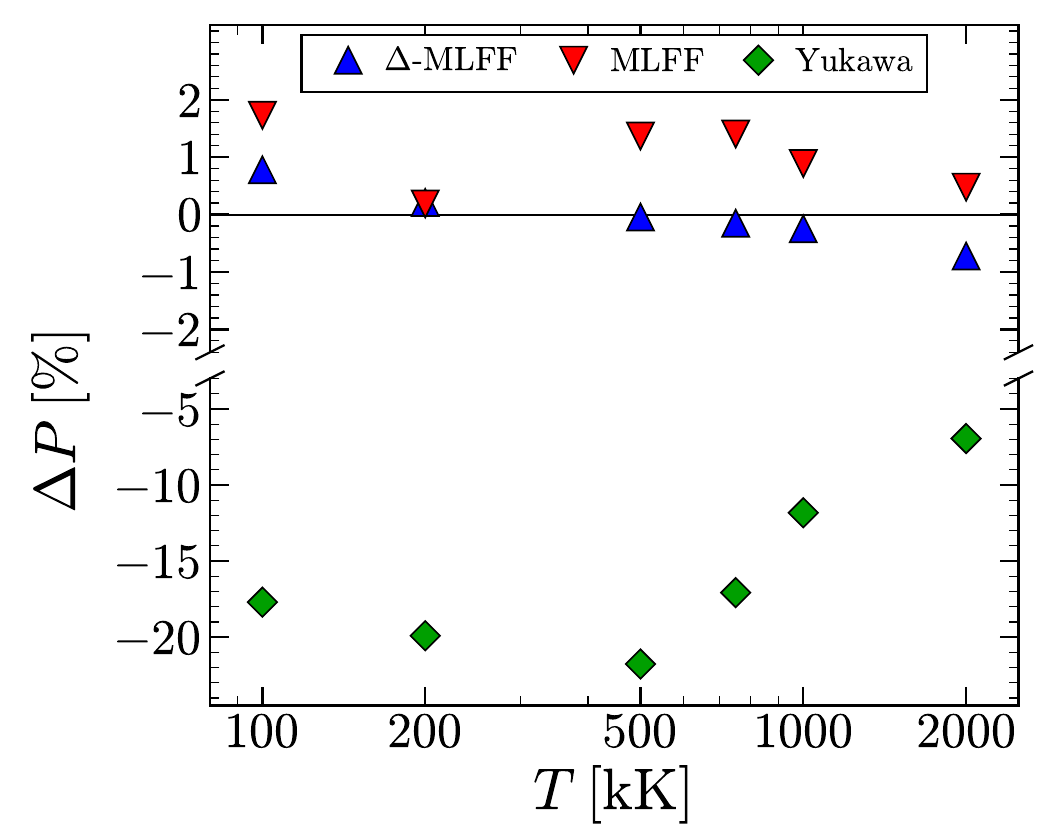}
    \caption{Relative deviation (\%) of the total pressure from DFT~\cite{bethkenhagen2023properties} for carbon at 10~g/cm$^3$ as a function of temperature, obtained using $\Delta$-MLFF, MLFF, and the Yukawa model. The statistical errors are smaller than the symbols.}
    \label{fig:C_P}
\end{figure}

In Fig.~\ref{fig:C_P}, we compare the total pressure of carbon at 10~g/cm$^3$ with the DFT equation of state of Ref.~\onlinecite{bethkenhagen2023properties}. We observe that $\Delta$-MLFF agrees with DFT to within 0.8\% at all six temperatures (deviations of $-0.72$ to $+0.78$\%, mean absolute deviation 0.36\%). In contrast, MLFF overestimates the pressure at every temperature, by 0.18--1.73\% (mean absolute deviation 1.01\%, more than twice that of $\Delta$-MLFF). Since the statistical errors of both the present and the reference pressures are below 0.05\%, these deviations are systematic. They are, however, comparable to the discretization error of up to 1.5\% in the reference pressures~\cite{bethkenhagen2023properties}, and may also reflect the differences in cell size and pseudopotential noted above for $g(r)$.

We find that the standalone Yukawa model underestimates the pressure by 7--22\%. The Yukawa pair potential, fitted to a few warm-start configurations, misses the part of the pressure that has no noticeable dependence on the ion positions, such as that of the nearly free electrons, since this part exerts no force on the ions. In $\Delta$-MLFF, this part is supplied by the machine-learned correction: along the $\Delta$-MLFF trajectories at 500 kK and 2 MK, the difference between the $\Delta$-MLFF and Yukawa pressures on the same configurations is nearly constant, at $3406 \pm 24$ and $4755 \pm 99$~GPa, respectively.

In summary, $\Delta$-MLFF agrees best with DFT (mean absolute deviation 0.36\%), followed by MLFF (1.01\%) and the standalone Yukawa model (15.87\%). The only exception is 200 kK, where $\Delta$-MLFF and MLFF are nearly equally close ($+0.21$ and $+0.18$\%, respectively).

\paragraph{Diffusion coefficient and viscosity}

\begin{figure*}[htbp]
    \centering
    \subfloat[]{\includegraphics[width=0.32\textwidth]{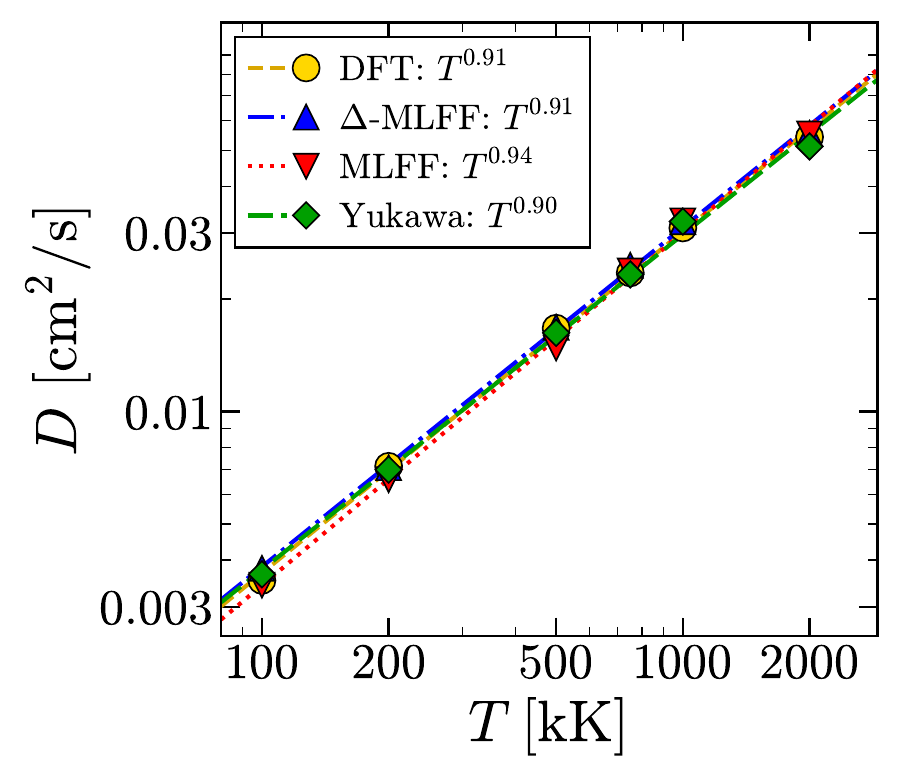}\label{fig:C_D}}\hfill
    \subfloat[]{\includegraphics[width=0.32\textwidth]{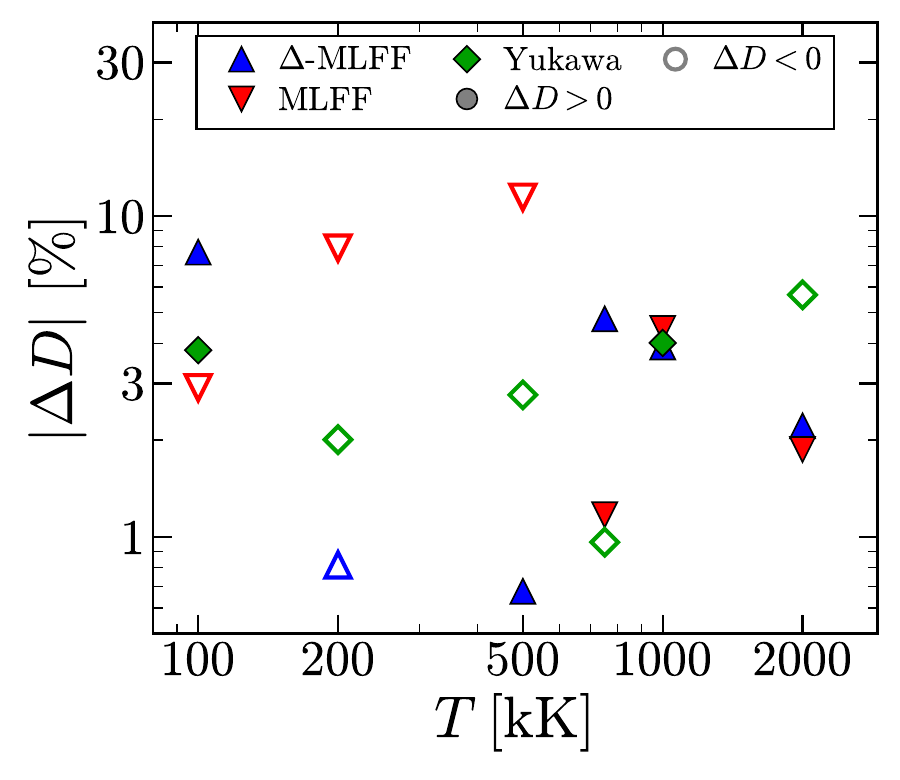}\label{fig:C_D_dev}}\hfill
    \subfloat[]{\includegraphics[width=0.32\textwidth]{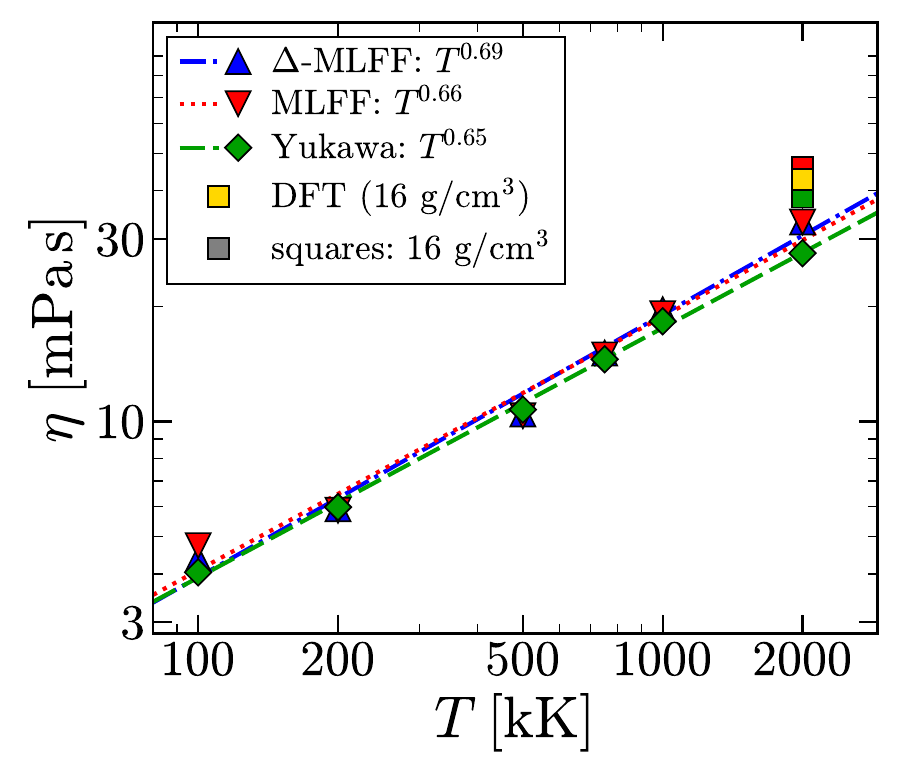}\label{fig:C_eta}}
    \caption{Self-diffusion coefficient $D$ and shear viscosity $\eta$ for carbon calculated using $\Delta$-MLFF, MLFF, and the Yukawa model, compared with DFT~\cite{bethkenhagen2023properties}. (a) $D$ as a function of temperature at 10~g/cm$^3$. (b) Relative deviation (\%) of $D$ from DFT. (c) $\eta$ as a function of temperature at 10~g/cm$^3$, together with the values at 16~g/cm$^3$ and 2 MK.}
    \label{fig:C_transport}
\end{figure*}

In Fig.~\ref{fig:C_transport}, we compare the self-diffusion coefficient and shear viscosity of carbon with DFT~\cite{bethkenhagen2023properties}. Along the 10~g/cm$^3$ isochore, we observe that $D$ follows a power law in $T$ for all methods, with an exponent of 0.91 for both DFT and $\Delta$-MLFF, 0.94 for MLFF, and 0.90 for the standalone Yukawa model (Fig.~\ref{fig:C_transport}(a)). The DFT values have statistical errors of 1.1--6.1\%, owing to the short 64-atom trajectories, which are comparable to the deviations of all three methods (Fig.~\ref{fig:C_transport}(b)). In particular, $\Delta$-MLFF deviates from DFT by $-0.8$ to $+7.7$\% (mean absolute deviation 3.3\%), agreeing with DFT to within two combined standard errors at all temperatures except 100 kK. MLFF deviates by $-11.4$ to $+4.5$\% (mean absolute deviation 5.0\%), underestimating $D$ at 200 kK and 500 kK by 7.9 and 11.4\%, respectively, i.e., by more than two combined standard errors. The standalone Yukawa model has a mean absolute deviation of 3.2\%, similar to that of $\Delta$-MLFF; however, at 2 MK, where the DFT error is smallest (1.1\%), it underestimates $D$ by 5.7\%, i.e., by five combined standard errors. Indeed, normalized by the combined standard errors, the RMS deviation is 1.3 for $\Delta$-MLFF, 2.1 for MLFF, and 2.2 for the standalone Yukawa model.

DFT data for the viscosity $\eta$ is available only at 16~g/cm$^3$ and 2 MK, where $\eta^{\rm DFT} = 42.8$~mPa\,s~\cite{bethkenhagen2023properties} (Fig.~\ref{fig:C_transport}(c)). At this state, we find that $\Delta$-MLFF agrees with DFT to within 1.4\% ($42.2 \pm 1.8$~mPa\,s), whereas MLFF is 7.5\% higher ($46.0 \pm 2.1$~mPa\,s) and the standalone Yukawa model is 10\% lower ($38.5 \pm 1.7$~mPa\,s). Along the 10~g/cm$^3$ isochore, $\eta$ follows a power law with exponents of 0.69, 0.66, and 0.65 for $\Delta$-MLFF, MLFF, and the standalone Yukawa model, respectively. We also observe that MLFF agrees with $\Delta$-MLFF to within 8\%, with the largest difference at 100 kK, whereas the standalone Yukawa model differs from $\Delta$-MLFF by $-17$ to $+3$\%, with the largest difference at 2 MK.

In summary, $\Delta$-MLFF is closest to DFT for both the diffusion coefficient $D$ and the viscosity $\eta$. For $D$, the standalone Yukawa model has a similar mean absolute deviation but deviates more relative to the DFT errors, while MLFF deviates the most. For $\eta$, $\Delta$-MLFF is followed by MLFF and then by the Yukawa model.

Overall, $\Delta$-MLFF is the most accurate of the three methods for all the properties considered. It has the lowest errors in the energy, forces, and stresses at all $\rho$--$T$ states, reproduces the DFT radial distribution function to within the spread between different DFT codes, agrees with the DFT pressure to within 0.8\%, and reproduces the DFT diffusion coefficient as well as the DFT viscosity at 16~g/cm$^3$ to within 1.4\%. The other two methods each fail in a different way. MLFF has force errors of 36--78\% and shear-stress errors of up to 188\%, and allows the ions to approach each other unphysically closely. Nevertheless, its pressure, diffusion coefficient, and viscosity deviate from DFT by only up to 1.7\%, 11\%, and 7.5\%, respectively, so that averaged properties alone do not reveal these errors. The standalone Yukawa model yields the correct short-range structure and a diffusion coefficient close to DFT on average, but misses 7--22\% of the pressure and underestimates the viscosity by 10\%. The Yukawa model thus supplies the short-range repulsion that MLFF lacks, while the machine-learned correction supplies the pressure and the remaining interactions that the Yukawa model lacks. Indeed, it is their combination in $\Delta$-MLFF that makes it accurate across all the properties.
\subsubsection{\label{subsubsec:performance}Performance}

\begin{figure}[htbp]
    \centering
    \includegraphics[width=0.8\linewidth]{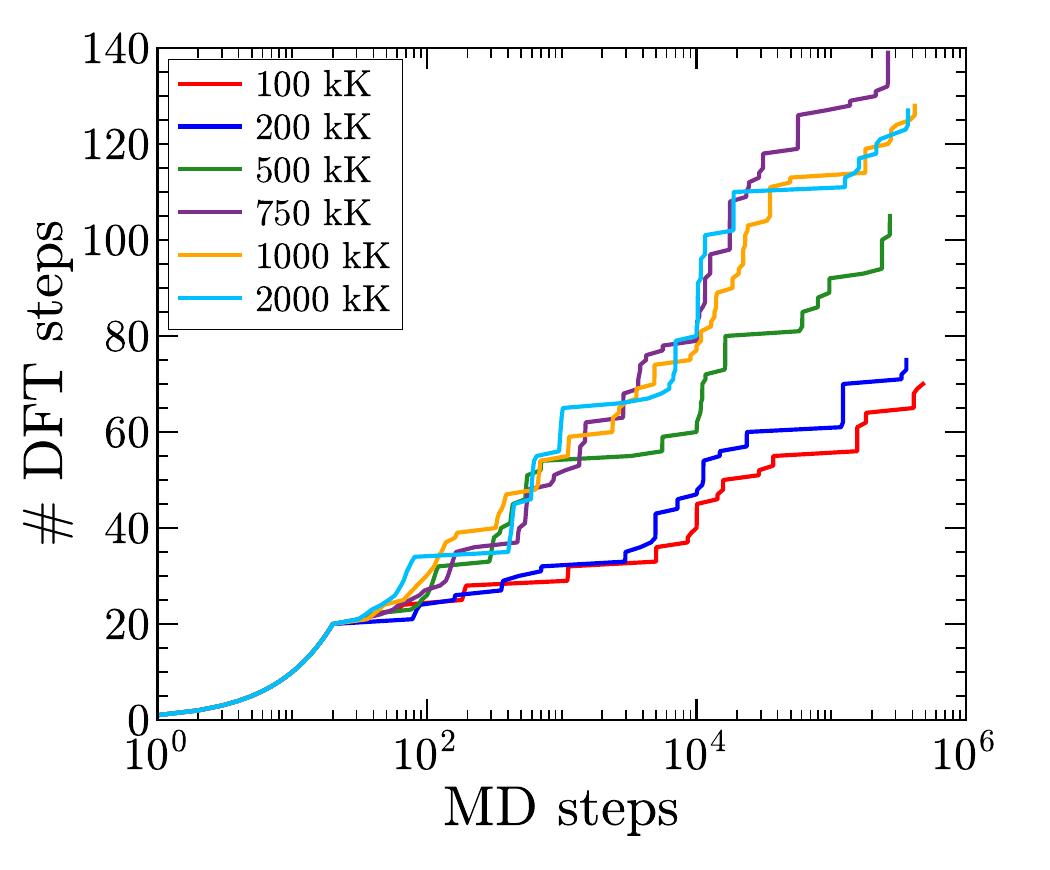}
    \caption{Cumulative number of DFT steps as a function of MD step during the on-the-fly $\Delta$-MLFF MD simulations for carbon at 10~g/cm$^3$. The first $N_1 = 5$ steps are used to fit the parameters of the Yukawa model, and the next $N_2 = 15$ steps to train the initial $\Delta$-MLFF model.}
    \label{fig:DFT_step_call}
\end{figure}

We next study the performance of the $\Delta$-MLFF scheme during the on-the-fly MD. In Fig.~\ref{fig:DFT_step_call}, we present the cumulative number of DFT steps at each temperature. We observe that the scheme requires relatively few DFT steps, 68--138, i.e., at most 0.028\% of the MD steps. Notably, this number is comparable to that of the standard MLFF scheme, 45--128, even though the $\Delta$-MLFF scheme is significantly more accurate, as shown in the previous section. Most of the DFT steps (69--86\%) occur within the first 50\,000 MD steps, after which they become rare but continue throughout the simulation.

In Table~\ref{tab:timing_comparison}, we list the resulting CPU times and speedups of the $\Delta$-MLFF scheme. We find that the speedups relative to AIMD are 678--2570. At 100 kK, the DFT steps of the 200-atom cell account for over 90\% of the total CPU time, whereas at the higher temperatures the DFT steps account for 28--67\%. With the same ML hyperparameters, the standard MLFF scheme requires a comparable number of DFT steps and has a comparable cost per ML step, and hence a similar speedup. A detailed comparison at equal accuracy is not possible, however, since the standard MLFF scheme cannot reach the accuracy of $\Delta$-MLFF. The $\Delta$-MLFF scheme thus provides 5--26 times lower force errors, and accuracy in all the properties discussed above, at a cost comparable to that of the standard MLFF scheme.

\begin{table}[htbp]
\caption{Performance of the $\Delta$-MLFF scheme for carbon at 10~g/cm$^3$. The $\Delta$-MLFF and DFT times per step are averaged over the respective steps of the 500\,000-step MD simulation. The number of DFT steps comprises the $N_1 = 5$ steps for fitting the Yukawa model, the $N_2 = 15$ steps for training the initial $\Delta$-MLFF model, and the subsequent on-the-fly DFT steps. The speedup is the ratio of the time an equivalent AIMD simulation, with DFT at every step, would take to the time of the $\Delta$-MLFF simulation.}
\label{tab:timing_comparison}
\begin{ruledtabular}
\begin{tabular}{rcccc}
 & & \multicolumn{2}{c}{Time (CPU s)} & \\
\cline{3-4}
$T$ (K) & \#DFT steps & $\Delta$-MLFF & DFT & Speedup \\
\hline
100 000   & 68  & 37.0 & 27 648 & 678  \\
200 000   & 74  & 22.8 & 86 400 & 2427 \\
500 000   & 104 & 11.3 & 62 400 & 2570 \\
750 000   & 138 & 7.8  & 57 000 & 2418 \\
1 000 000 & 127 & 5.9  & 27 360 & 2131 \\
2 000 000 & 126 & 2.3  & 3456   & 1089 \\
\end{tabular}
\end{ruledtabular}
\end{table}

Since the $\Delta$-MLFF scheme is significantly more accurate than the standard MLFF scheme, its excess accuracy can be traded for speed, making it faster than the standard MLFF scheme at the same accuracy. We test this at 2 MK and 10~g/cm$^3$, where the standard MLFF scheme requires 128 DFT steps, takes 3.2 CPU s per MD step, and has mean errors of 49\% in the forces, 2.2\% in the diagonal stress, and 121\% in the off-diagonal stress. We find that raising the scaling factor for $s_{\rm tol}$ from 1.001 to 1.01 alone reduces the number of DFT steps in the 500\,000-step simulation from 126 to 35,  lowering the CPU time per MD step from 3.2 to 2.5 s, while the mean energy error rises only from 3.0 to 4.0~mHa/atom, the mean force error from 1.9 to 3.3\%, and the mean off-diagonal-stress error from 4.6 to 7.3\%, with the mean diagonal-stress error remaining at about 0.6\%. The $\Delta$-MLFF scheme thus becomes 22\% faster than the standard MLFF scheme, while its force and off-diagonal-stress errors remain more than an order of magnitude smaller.  We also consider two other ways of trading accuracy for speed. First, coarsening the DFT calculations (SQ quadrature order $30 \to 20$, grid spacing $0.13 \to 0.20$ bohr) introduces a discretization error of about 1.6\% in the diagonal stress and reduces the cost of a DFT step from 3456 to 384 CPU s; with 135 DFT steps, the CPU time per MD step drops to 2.4 s. Second, coarsening the machine-learned model (SOAP cutoff radius $10 \to 5$ bohr, scaling factor for $s_{\rm tol}$ $1.001 \to 1.01$) makes the ML step faster; with 166 DFT steps, the CPU time per MD step drops to 2.7 s, at a diagonal-stress error of about 2.2\%. In all three cases, we observe that the $\Delta$-MLFF scheme is faster than the standard MLFF scheme (2.4--2.7 versus 3.2 CPU s per MD step) at better or comparable stress accuracy, while retaining the short-range repulsion of the Yukawa model.


\subsection{\label{subsec:Mg}Application: warm dense magnesium}

We now apply the $\Delta$-MLFF framework to warm dense magnesium (Mg). As a constituent of MgO and MgSiO$_3$, Mg is one of the fundamental building blocks of rocky planets, with Mg-bearing matter reaching warm dense conditions in planetary cores and during giant impacts \cite{gonzalez2020equation,soubiran2019magnesium}. Indeed, Mg has been the subject of first-principles equation-of-state studies \cite{gonzalez2020equation} as well as dynamic-compression experiments \cite{Beason2020}. Moreover, being inexpensive and readily fabricated into thin foils and layered targets, it represents a practical choice for high-energy-density experiments. The properties computed here are directly relevant to such experiments: the pressure determines the equation of state entering hydrodynamic simulations of the compression process, the diffusion coefficient and viscosity govern mixing and dissipation in the resulting plasma, and the dynamic structure factor is directly measurable via X-ray Thomson scattering diagnostics \cite{Glenzer2009}.

\begin{figure}[htbp]
    \centering
    \includegraphics[width=0.8\linewidth]{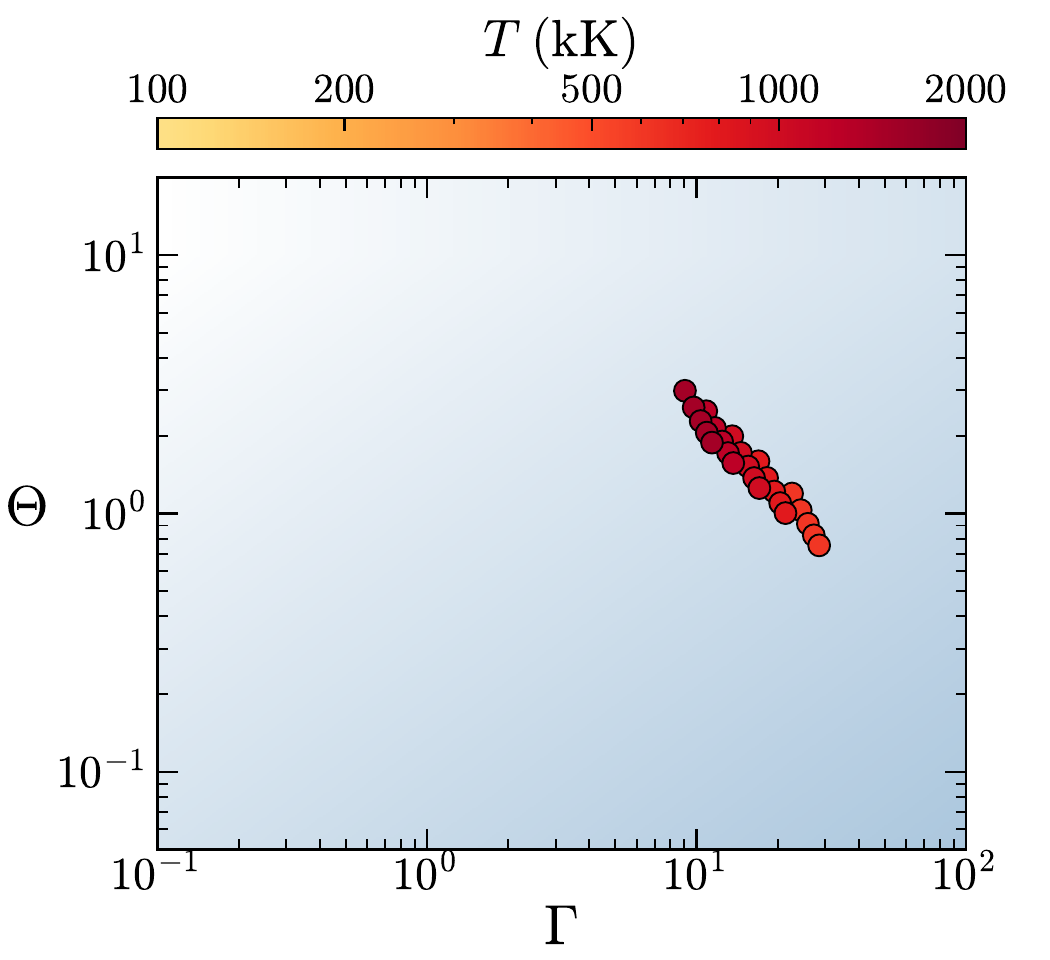}
    \caption{Electron degeneracy parameter ($\Theta$) and ionic coupling parameter ($\Gamma$) for Mg at the densities and temperatures considered in this work.}
    \label{fig:theta_gamma_Mg}
\end{figure}

We consider Mg at densities of $\rho = 5.22$, 6.525, 7.83, 9.135, and 10.44~g/cm$^3$, corresponding to compression ratios of $\rho/\rho_0 = 3$, 3.75, 4.5, 5.25, and 6 with respect to the ambient density $\rho_0 = 1.74$~g/cm$^3$, and temperatures of $T = 600$~kK, 800~kK, 1~MK, 1.25~MK, and 1.5~MK, for a total of 25 $\rho$-$T$ states. These conditions lie in the WDM regime, with plasma parameters of $\Theta = 0.75$--$3.0$ and $\Gamma = 9.1$--$28.5$ (Fig.~\ref{fig:theta_gamma_Mg}), indicative of partially degenerate electrons and strongly coupled ions, which motivates the need for ab initio accuracy \cite{stanek2024review}. The average ionization is $Z^* = 9.97$--$9.99$ across all the $\rho$-$T$ states considered.

The Mg simulations use cells containing $N = 119$, 106, 95, 85, and 77 atoms at temperatures of 600~kK, 800~kK, 1~MK, 1.25~MK, and 1.5~MK, respectively, at all densities. The time step ranges from 0.024~fs to 0.049~fs, chosen relative to the reference value of $\Delta t = 0.04$~fs at $\rho = 4.31$~g/cm$^3$ and $T = 1$~MK, which has been found to be sufficient \cite{suryanarayana2025accuracy}. Ten electrons of Mg are treated as valence, with an ONCV pseudopotential having a cutoff radius of 0.50~bohr, two projectors each for the $s$ and $p$ channels, and one projector for the $d$ channel. The electronic structure is computed using the SQ method at all the $\rho$-$T$ states. The grid spacing is 0.13~bohr, while the quadrature order ranges from 60 to 30 and the truncation radius from 1.5~bohr to 0.75~bohr, with the first value of each range corresponding to the lowest temperature and the second to the highest.

\paragraph{Dynamic structure factor}

\begin{figure*}[htbp]
    \centering
    \subfloat[]{\includegraphics[width=0.38\textwidth]{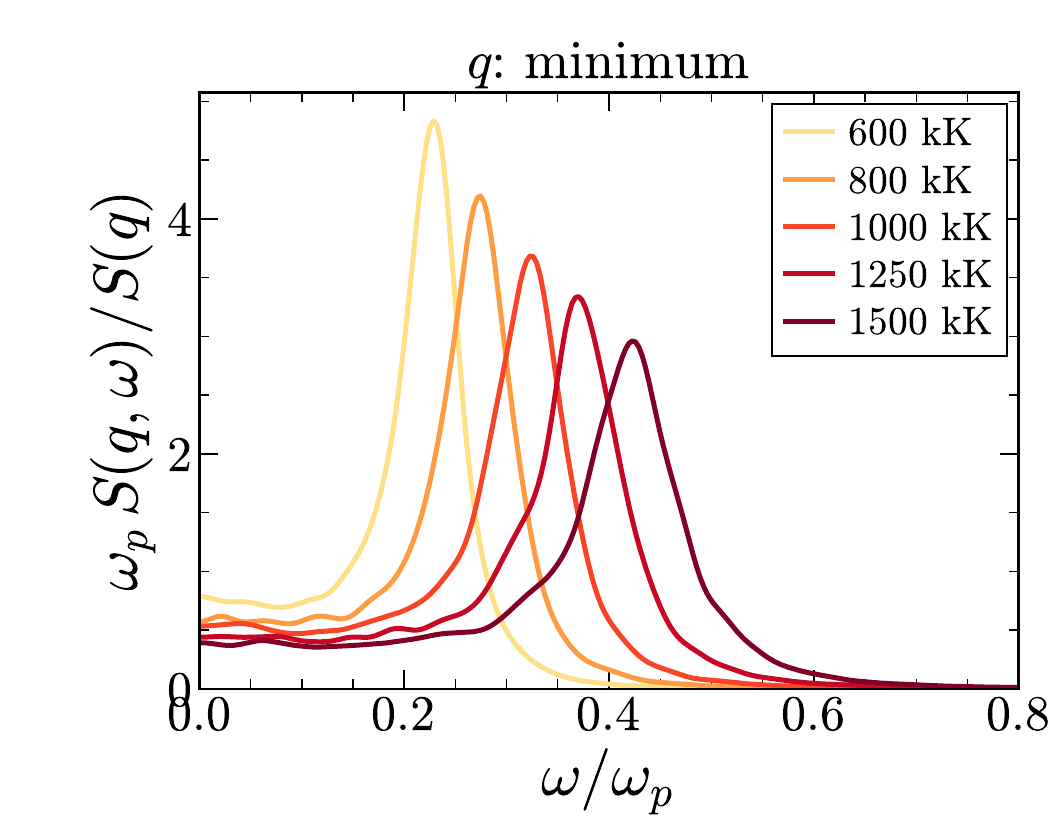}\label{fig:Mg_skw_a}}
    \subfloat[]{\includegraphics[width=0.38\textwidth]{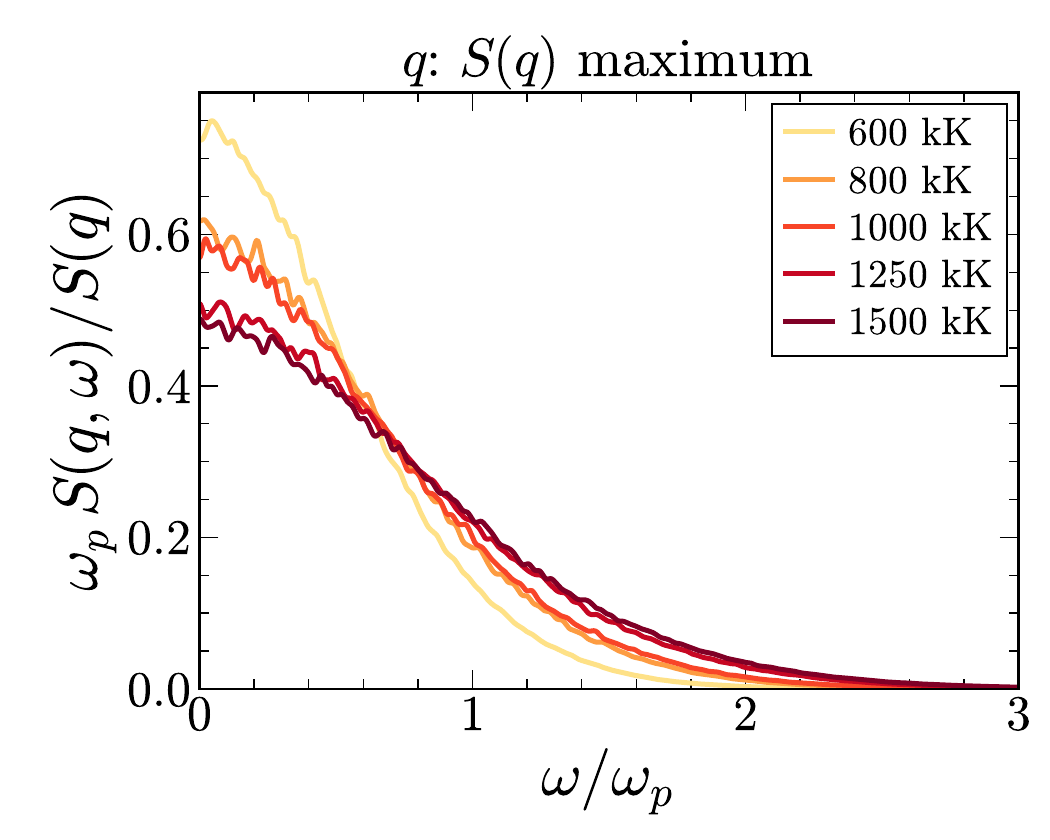}\label{fig:Mg_skw_b}}\\
    \subfloat[]{\includegraphics[width=0.38\textwidth]{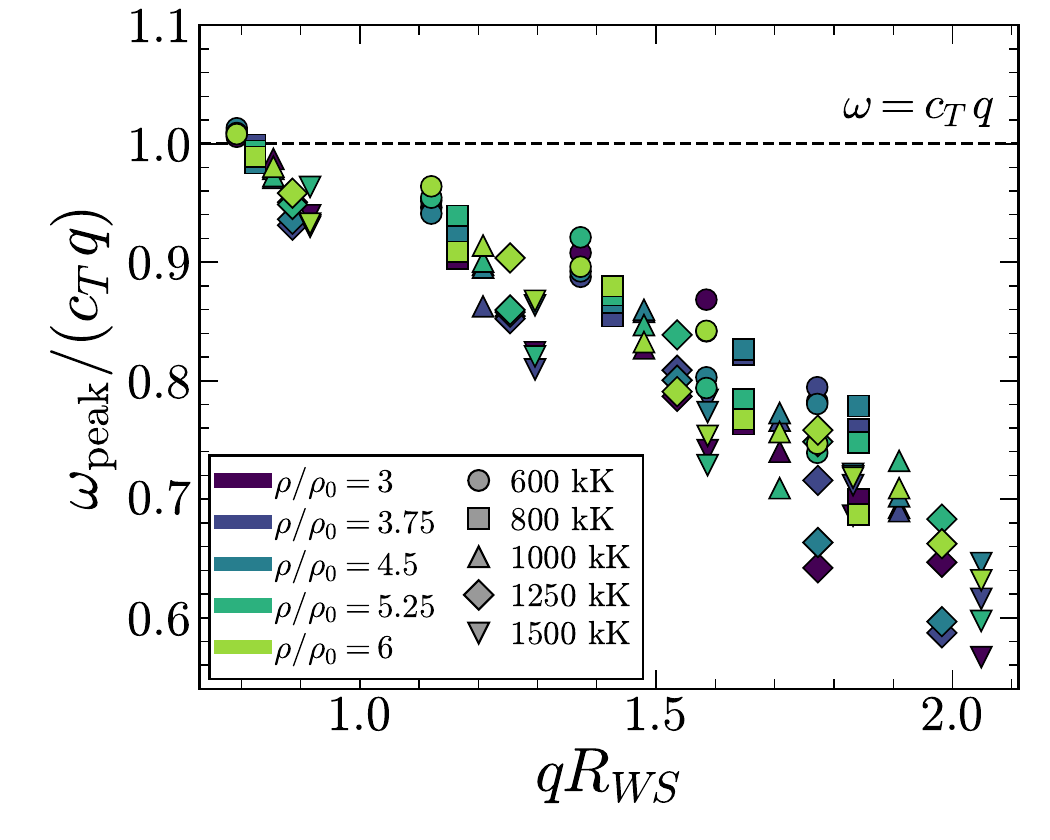}\label{fig:Mg_skw_c}}
    \subfloat[]{\includegraphics[width=0.38\textwidth]{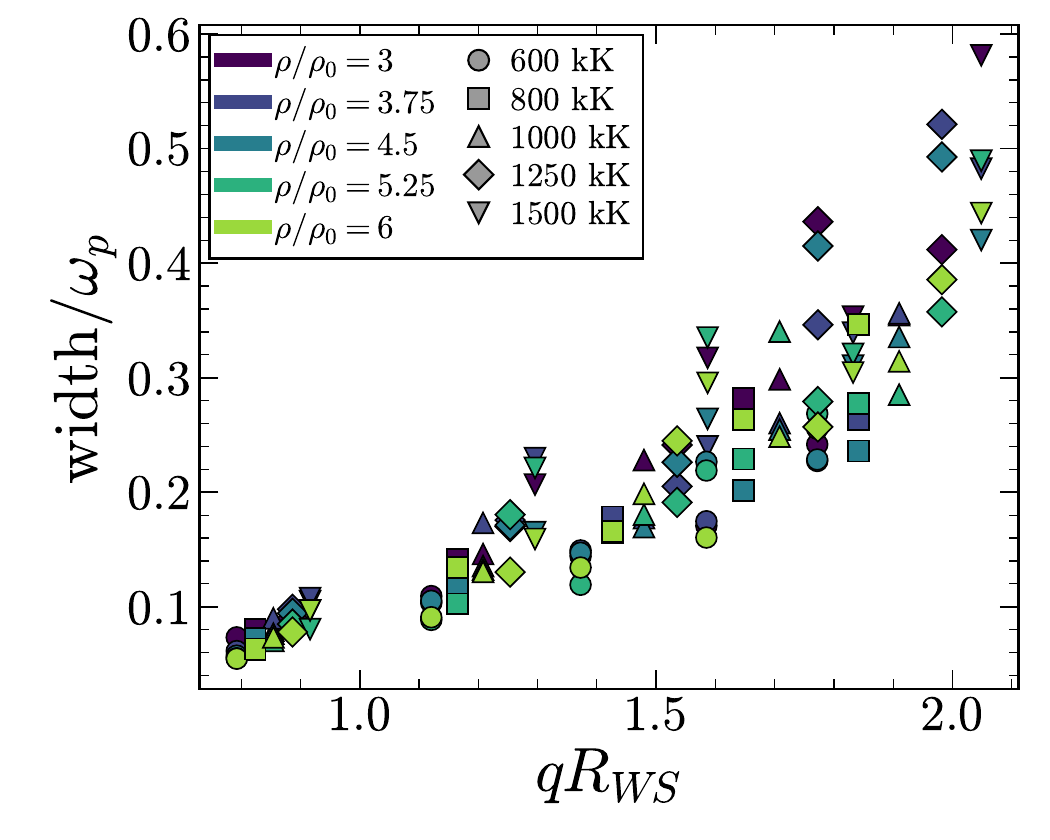}\label{fig:Mg_skw_d}}
    \caption{Ionic dynamic structure factor $S(q,\omega)$ for Mg calculated using the $\Delta$-MLFF scheme. (a) $\omega_p S(q,\omega)/S(q)$ at the smallest wavenumber $q$ of the simulation cell, where $S(q)$ is the static structure factor and $\omega_p$ is the ion plasma frequency. (b) The same quantity at the wavenumber $q$ at which $S(q)$ is maximum. (c) Phase velocity $\omega_{\rm peak}/q$ of the ion-acoustic peak, with $\omega_{\rm peak}$ the frequency of its maximum, relative to the isothermal sound speed $c_T$. (d) Width of the ion-acoustic peak, defined as twice the half width at half maximum on its high-frequency side, in units of $\omega_p$. Panels (a) and (b) are at a compression ratio of $\rho/\rho_0 = 4.5$ and are shown as functions of $\omega/\omega_p$. Panels (c) and (d) are for the five smallest wavenumbers $q$ of the simulation cell and are shown as functions of $qR_{WS}$, where $R_{WS}$ is the Wigner-Seitz ion-sphere radius.}
    \label{fig:Mg_skw}
\end{figure*}

In Fig.~\ref{fig:Mg_skw}, we present the dynamic structure factor $S(q,\omega)$, which characterizes the collective ion dynamics. In panels (a) and (b), we display $S(q,\omega)$, normalized by the static structure factor $S(q)$, at the smallest wavenumber accessible in the simulation cell and at the wavenumber at which $S(q)$ attains its maximum, respectively, for a compression ratio of $\rho/\rho_0 = 4.5$, while in panels (c) and (d), we summarize the ion-acoustic peak for the five smallest wavenumbers at all the $\rho$-$T$ states. We express frequencies in units of the ion plasma frequency $\omega_p = (4\pi n_i {Z^*}^2/M)^{1/2}$, where $Z^* \approx 10$ is the DFT mean ionization, corresponding to a fully ionized L shell at all conditions considered, and wavenumbers in units of the inverse Wigner-Seitz ion-sphere radius $R_{WS} = (3/4\pi n_i)^{1/3}$, allowing results at different densities to be compared directly. We observe that at the smallest wavenumber accessible in the simulation cell, $qR_{WS} \approx 0.8$--$0.9$, determined solely by the number of ions $N = 77$--$119$ used at the different temperatures, $S(q,\omega)$ exhibits a well-defined ion-acoustic peak, indicating that density fluctuations propagate as collective waves. In particular, this peak shifts from $0.23\,\omega_p$ to $0.42\,\omega_p$ between the lowest and highest temperatures, while broadening with increasing temperature (Fig.~\ref{fig:Mg_skw}(a)). At the maximum of $S(q)$, $qR_{WS} \approx 5.0$--$5.7$, where the wavelength is comparable to the interparticle spacing, we find that no well-defined ion-acoustic peak remains, with only a central peak at $\omega = 0$ being observed, indicative of local relaxation of the ions (Fig.~\ref{fig:Mg_skw}(b)). The half width of this central peak is $0.85$--$0.95$ times that of a free-particle gas, $\sqrt{2\ln 2}\, q\, (k_B T/M)^{1/2}$. Notably, only $2$--$3\%$ of this narrowing is accounted for by the de~Gennes factor $S(q)^{-1/2}$ \cite{degennes1959liquid, hansen2013theory}, since $S(q) = 1.04$--$1.07$ at its maximum, with the remainder reflecting the collisional slowing of local density relaxation in the strongly coupled ion liquid.

We further characterize the ion-acoustic peak by its phase velocity $\omega_{\rm peak}/q$, with $\omega_{\rm peak}$ denoting the frequency of its maximum, and by its width, defined as twice the half width at half maximum on its high-frequency side, which provides a measure of the damping of the mode. In Figs.~\ref{fig:Mg_skw}(c) and \ref{fig:Mg_skw}(d), we present both quantities. In the above units, we observe that the results at different densities nearly coincide, with the trends governed by the temperature. At the smallest wavenumber, we find that the phase velocity agrees to within 7\% with the isothermal sound speed $c_T = [(\partial P/\partial \rho)_T]^{1/2} = 35$--$63$ km/s obtained from the $\Delta$-MLFF equation of state, which is related to the static structure factor through the compressibility sum rule $S(q \to 0) = k_B T/(M c_T^2)$, the extrapolated ionic $S(q \to 0)$ agreeing with this value to within $10\%$. The phase velocity lies systematically below $c_T$ ($0.93$--$1.01\,c_T$ in 20 of the 25 $\rho$-$T$ states), indicating that dispersion has already set in at the smallest wavenumber. With increasing wavenumber, the phase velocity decreases to $0.57$--$0.79\,c_T$ at $qR_{WS} \approx 1.8$--$2$. Part of this decrease is caused by the growing central peak, which shifts the maximum of $S(q,\omega)$ toward $\omega = 0$. Indeed, the maximum of the longitudinal current correlation $C_L(q,\omega) = (\omega^2/q^2)\,S(q,\omega)$, which is free of this effect, gives $0.73$--$0.87\,c_T$ at the same wavenumbers, whereby about $70\%$ of the decrease is a property of the mode itself. Note that $\omega_{\rm peak}$ continues to increase with $q$, i.e., the negative dispersion refers to the phase velocity falling below $c_T$, and not to $d\omega_{\rm peak}/dq < 0$. Over the same range, the peak width, shown in units of $\omega_p$, grows approximately as $q^2$, reaching $0.5$--$1$ times the peak frequency. The same flattening of the peak frequency and growth of the width with wavenumber have been observed in simulations of the Yukawa one-component plasma~\cite{mithen2011density} for $\Gamma = 1$--$175$, which encompasses the range $\Gamma = 9.1$--$28.5$ of the present study, as well as in DFT simulations of warm dense aluminum~\cite{ruter2014ab}. Since the MD is performed within the Born--Oppenheimer approximation, electron--ion collisional damping is not included, which is expected to strengthen the central peak and enhance the damping of the ion-acoustic mode~\cite{mabey2017strong}. The widths reported here can therefore be considered to be lower bounds.

\begin{figure*}[htbp]
    \centering
    \subfloat[]{\includegraphics[width=0.33\textwidth]{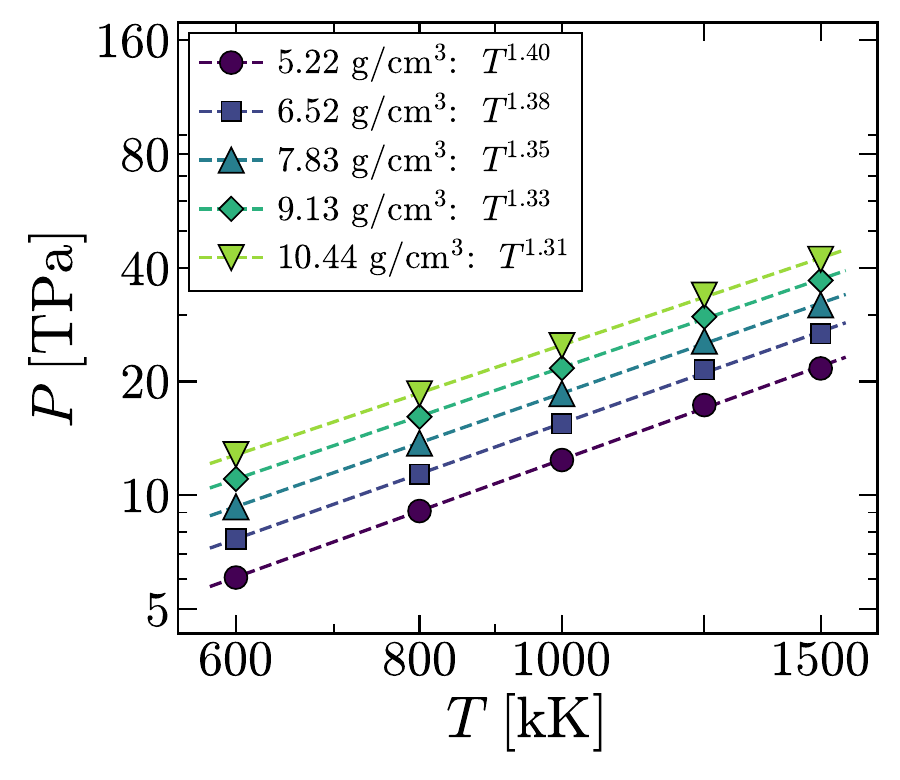}\label{fig:Mg_P_T}}\hfill
    \subfloat[]{\includegraphics[width=0.33\textwidth]{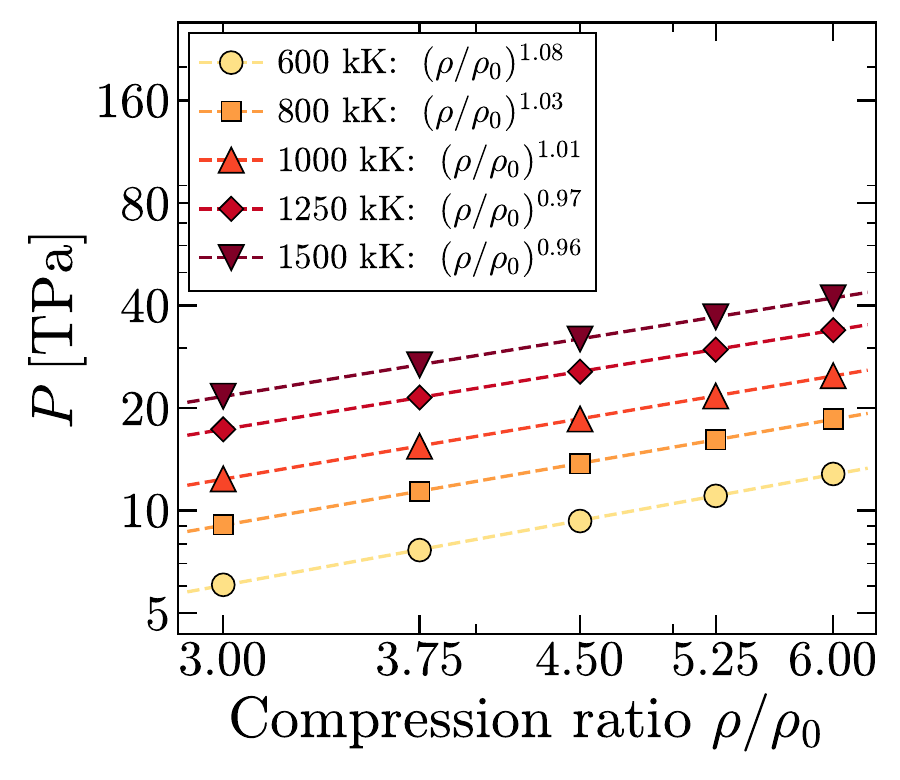}\label{fig:Mg_P_CR}}\hfill
    \subfloat[]{\includegraphics[width=0.33\textwidth]{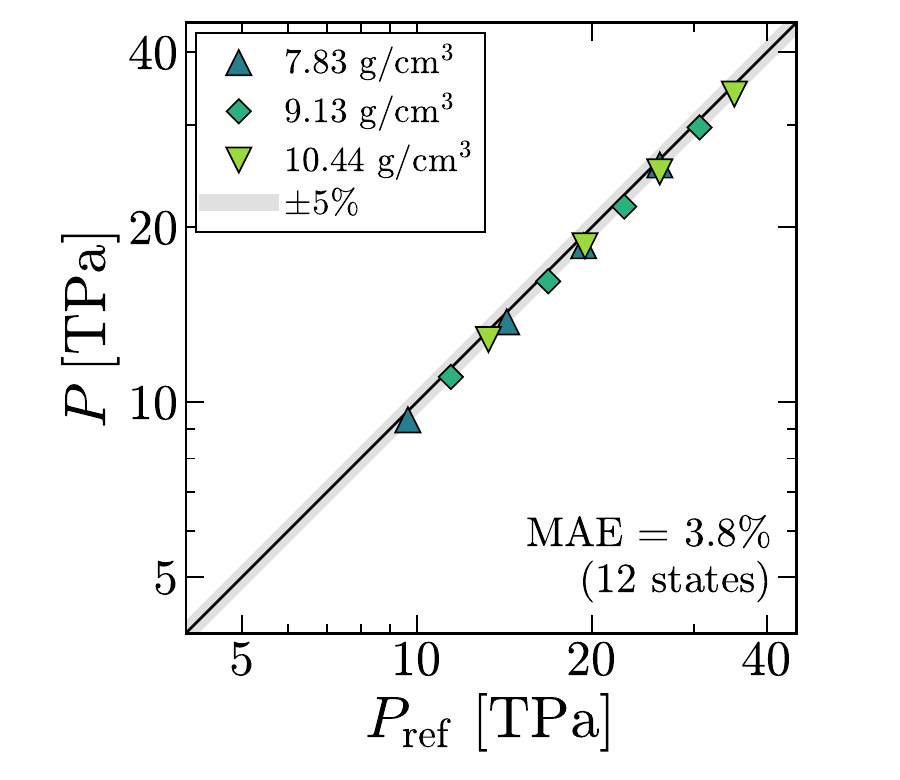}\label{fig:Mg_P_parity}}
    \caption{Total pressure $P$ for Mg calculated using the $\Delta$-MLFF scheme (a) as a function of temperature along isochores and (b) as a function of compression ratio along isotherms. (c) Comparison with the pressure $P_{\rm ref}$ obtained from the DFT equation of state of Ref.~\onlinecite{gonzalez2020equation}, interpolated to the 12 $\rho$-$T$ states of this work that lie within its density--temperature range. The statistical errors are smaller than the symbols.}
    \label{fig:Mg_pressure}
\end{figure*}

\paragraph{Pressure}
In Fig.~\ref{fig:Mg_pressure}, we present the total pressure for Mg along the isochores and isotherms. The pressure ranges from 6.1 to 42~TPa over the 25 $\rho$-$T$ states, with statistical errors below 0.01\%. We observe that along each isochore, the pressure increases faster than linearly with temperature, following a power law with exponent 1.31--1.40, whereas along each isotherm, it increases nearly linearly with density, following a power law with exponent 0.96--1.08. Notably, the pressure is 52--74\% of that for an ideal gas of ions and free electrons, $(1+Z^*)\,n_i k_B T$, approaching this value with increasing temperature. We therefore conclude that the interactions reduce the pressure, with a relative contribution that diminishes as the temperature increases.

To assess the accuracy of these pressures, we compare them with an independent DFT equation of state. At the 12 $\rho$-$T$ states that lie within the density--temperature range of the equation of state of Ref.~\onlinecite{gonzalez2020equation}, i.e., $\rho \geq 7.83$~g/cm$^3$ and $T \leq 1.25$~MK, we find that the $\Delta$-MLFF pressure is 2.3--4.9\% lower than the reference, with a mean deviation of 3.8\% (Fig.~\ref{fig:Mg_pressure}(c)). Since none of the reference states coincides with those considered here, we interpolate the reference pressure, with an estimated interpolation error of up to 2.2\%. The remaining difference is consistent with the differences in cell size, pseudopotential, and MD time step, as well as the error bars reported in Ref.~\onlinecite{gonzalez2020equation}.
\paragraph{Diffusion coefficient and viscosity}

\begin{figure*}[htbp]
    \centering
    \subfloat[\label{fig:Mg_D_T}]{%
        \includegraphics[width=0.325\linewidth]{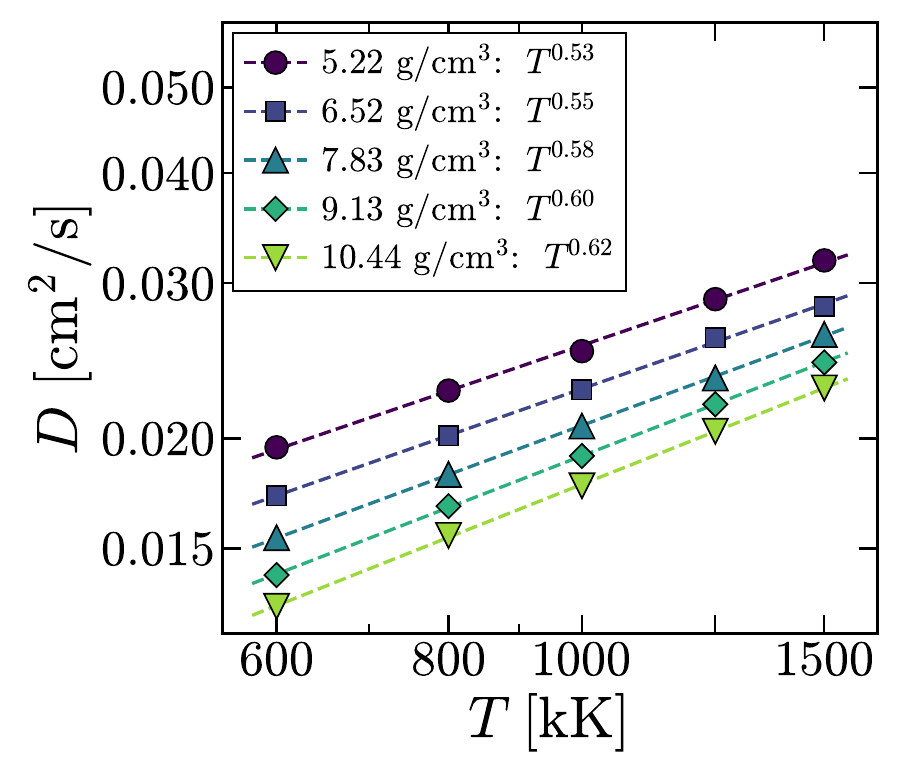}}
    \hfill
    \subfloat[\label{fig:Mg_eta_T}]{%
        \includegraphics[width=0.325\linewidth]{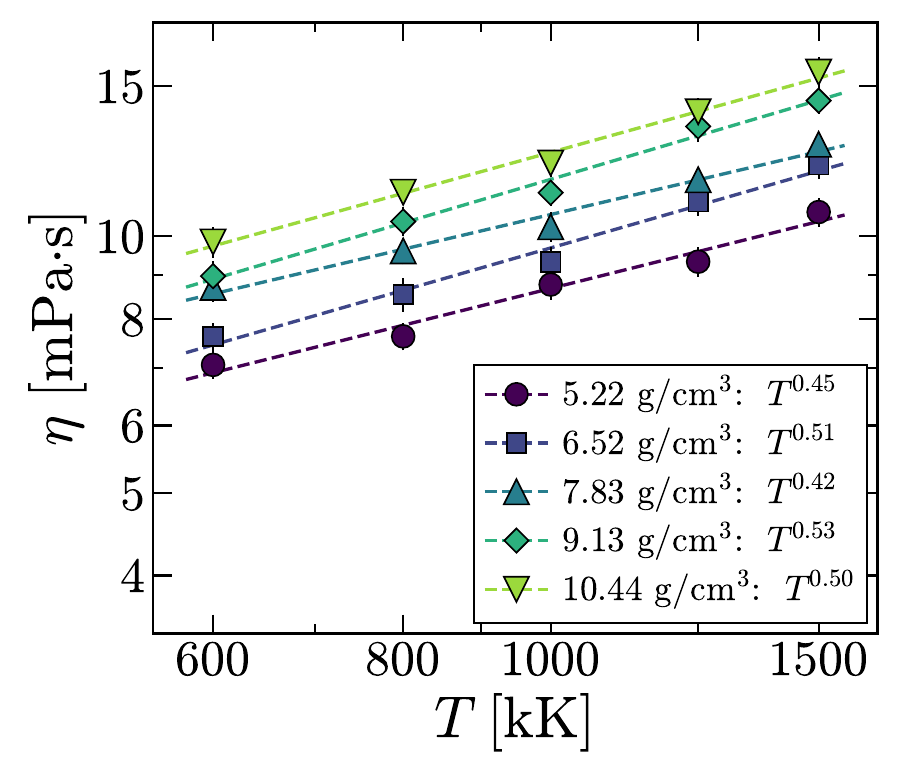}}
    \hfill
    \subfloat[\label{fig:Mg_alpha_T}]{%
        \includegraphics[width=0.325\linewidth]{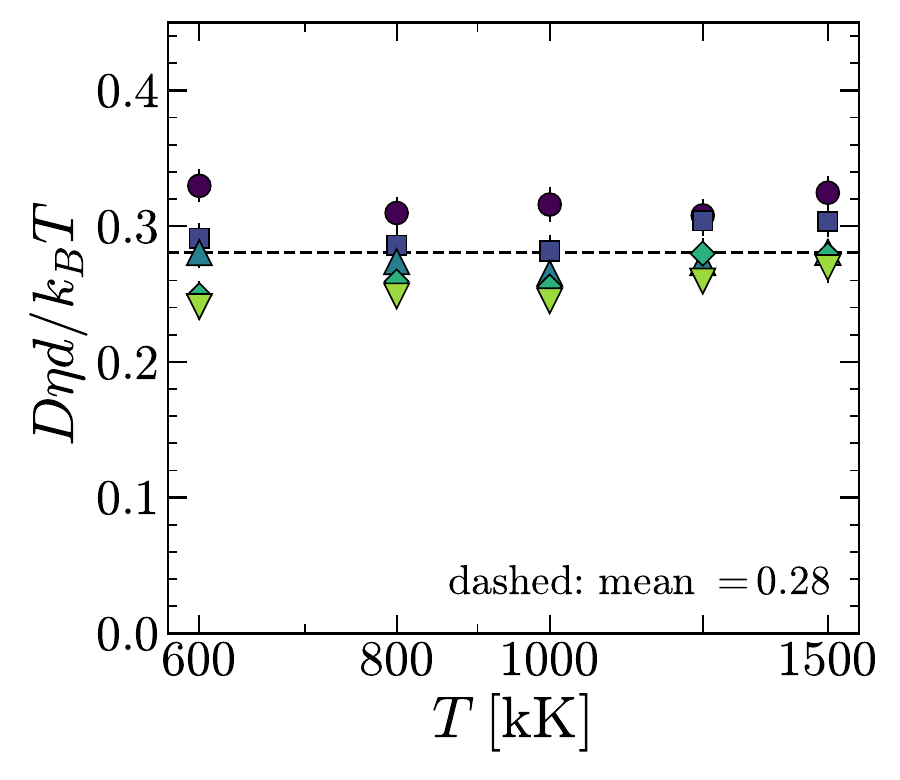}}

    \subfloat[\label{fig:Mg_D_CR}]{%
        \includegraphics[width=0.325\linewidth]{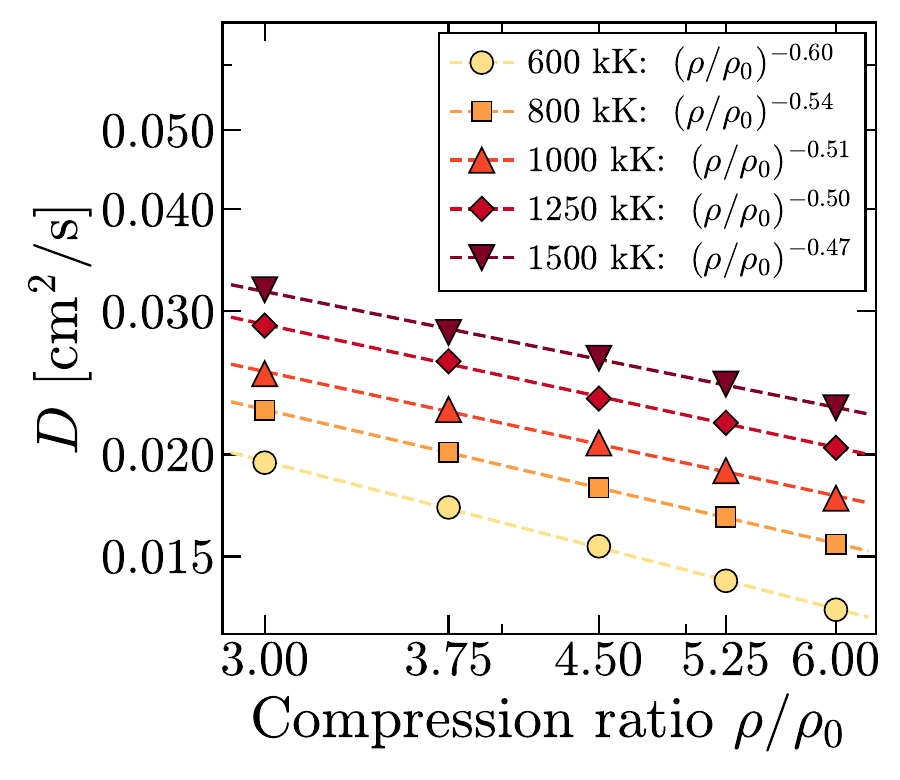}}
    \hfill
    \subfloat[\label{fig:Mg_eta_CR}]{%
        \includegraphics[width=0.325\linewidth]{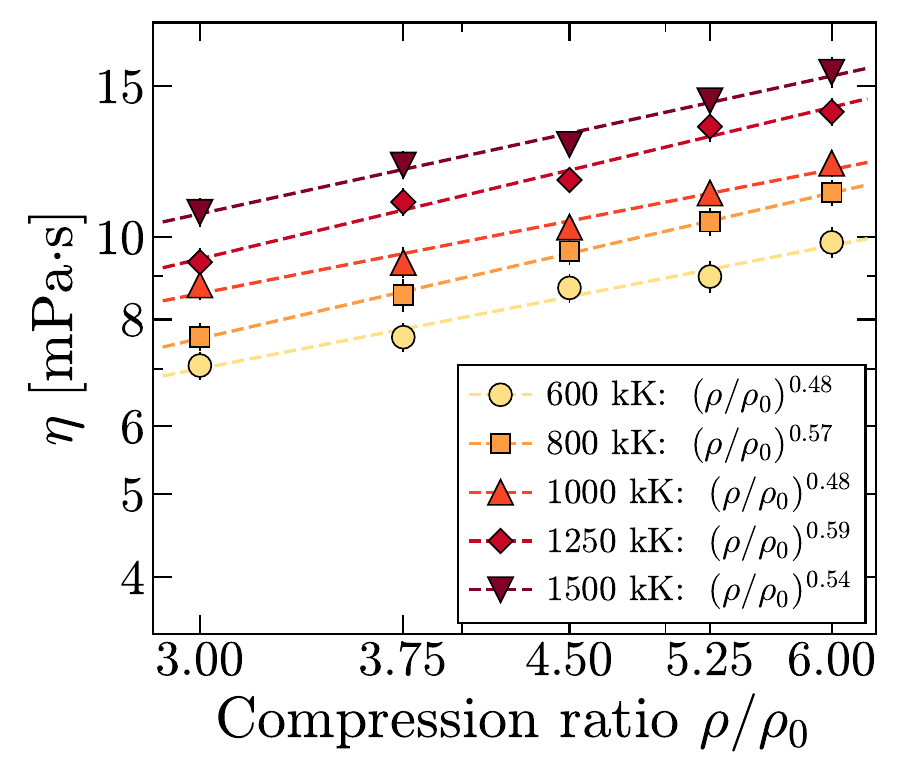}}
    \hfill
    \subfloat[\label{fig:Mg_alpha_CR}]{%
        \includegraphics[width=0.325\linewidth]{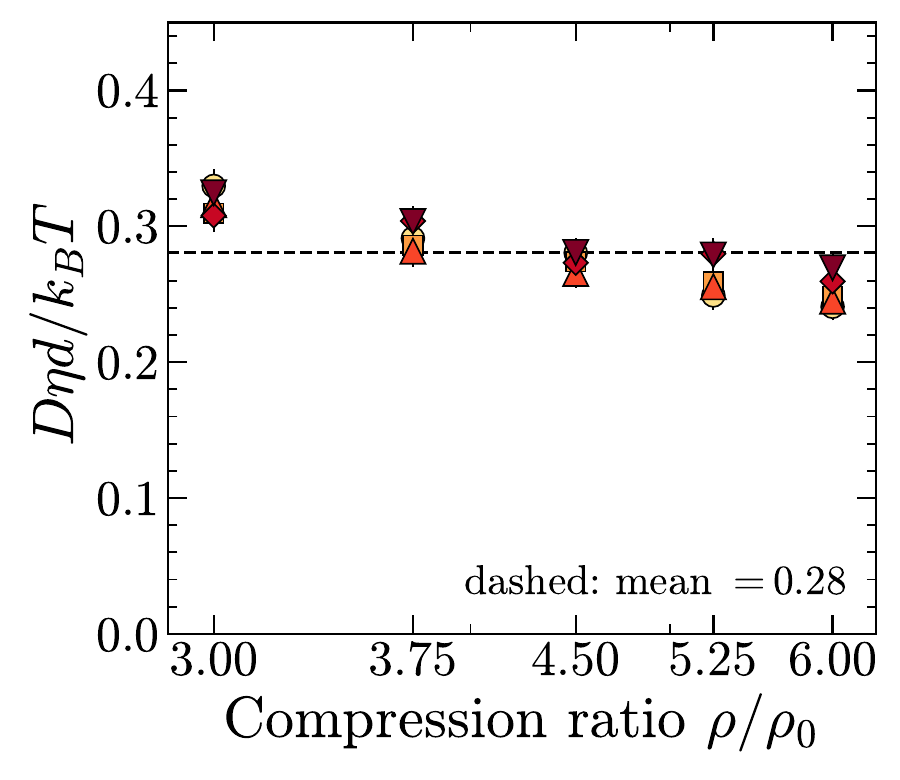}}
    \caption{Transport properties for Mg calculated using the $\Delta$-MLFF scheme: self-diffusion coefficient $D$, shear viscosity $\eta$, and Stokes--Einstein coefficient $\alpha_{\rm SE} = D\eta\,d/(k_BT)$, with $d = n_i^{-1/3}$ the mean interparticle distance. Panels (a)--(c) show the results as a function of temperature along isochores, and panels (d)--(f) as a function of compression ratio along isotherms.}
    \label{fig:Mg_transport}
\end{figure*}

In Fig.~\ref{fig:Mg_transport}, we present the self-diffusion coefficient $D$ and the shear viscosity $\eta$ for Mg along the isochores and isotherms. We observe that along each isochore, both quantities increase with temperature, following power laws with exponents of 0.53--0.62 for $D$ and 0.42--0.53 for $\eta$, whereas along each isotherm, $D$ decreases while $\eta$ increases with compression, following power laws with exponents of $-0.60$ to $-0.47$ for $D$ and 0.48 to 0.59 for $\eta$. The increase of $\eta$ with temperature indicates that these $\rho$-$T$ states, with $\Gamma = 9.1$--$28.5$, lie on the kinetic side of the viscosity minimum of strongly coupled Yukawa liquids~\cite{donko2008}.

The opposite dependence of $D$ and $\eta$ on density, together with their similar increase with temperature, suggests examining their product. In particular, the exponents in temperature sum to approximately one, while those in density nearly cancel, so that $D\eta$ scales approximately as $T$ and is nearly independent of density. This corresponds to the Stokes--Einstein relation, which for simple fluids can be written without a hydrodynamic radius as $\alpha_{\rm SE} = D\eta\,d/(k_BT) \approx {\rm const}$, with $d = n_i^{-1/3}$ denoting the mean interparticle distance~\cite{khrapak2019stokes}. Over the 25 $\rho$-$T$ states, we find that $\alpha_{\rm SE}$ has a mean of 0.28 with a standard deviation of 9\%, whereas $D$ and $\eta$ individually vary by factors of 2.5 and 2.2, respectively. Furthermore, we observe that along each isochore, $\alpha_{\rm SE}$ is nearly independent of temperature, while decreasing weakly with density, from 0.33 to 0.24 between $\rho/\rho_0 = 3$ and 6. We therefore conclude that the Stokes--Einstein relation provides an estimate of either transport coefficient from the other, to within about 10\% over the conditions considered here.


\section{\label{sec:conclusion}Concluding Remarks}

In this work, we have developed a framework for on-the-fly $\Delta$-MLFF MD simulations of WDM, wherein we have employed a Yukawa reference model that captures the dominant screened-Coulomb interactions between the ions, with its parameters fit on the fly to Kohn--Sham DFT data, in conjunction with an MLFF scheme based on the kernel method and Bayesian linear regression that learns the differences in the energies, atomic forces, and stresses between the Yukawa model and Kohn--Sham DFT. We have implemented this framework in the real-space electronic structure code SPARC, extending its on-the-fly MLFF capability. Choosing warm dense carbon as a representative example, we have verified the accuracy and performance of the formalism through comparisons with Kohn--Sham DFT results in the literature. In so doing, we have demonstrated that $\Delta$-MLFF has significantly lower errors in the energies, atomic forces, and stresses than the standard MLFF scheme at a comparable cost, and that AIMD simulations of WDM can be accelerated by up to three orders of magnitude while retaining ab initio accuracy. Furthermore, we have applied this framework to calculate the equation of state, structure, and transport properties of magnesium at densities of 5.22 to 10.44~g/cm$^3$ and temperatures of 600~kK to 1.5~MK. We have found that the pressure, self-diffusion coefficient, and shear viscosity obey power laws in temperature and density, that the dynamic structure factor exhibits a well-defined ion-acoustic mode whose negative dispersion is characteristic of strongly coupled liquids, and that the Stokes--Einstein relation holds to within about 10\% over the conditions considered. Overall, the proposed formalism promises to enable simulations of WDM with ab initio accuracy at a small fraction of the cost of AIMD.

The Bayesian error estimate that triggers the DFT calculations is inherited from the standard MLFF scheme, and the results for carbon indicate that it is conservative. Calibrating this estimate against the errors observed at the DFT calls, while also accounting for the error of the Yukawa model itself, could further reduce the number of DFT calculations required. In addition, the development of specialized kernels and descriptors for the learned correction, which is smoother and of smaller magnitude than the quantities themselves, is likely to improve the accuracy and transferability of the model. Finally, extending the current implementation to enable GPU acceleration of the key computational kernels \cite{sharma2023gpu, jing2025gpu} will further reduce the wall time for large MD simulations. These represent worthy subjects for future research.


\section*{Supplementary Material}
$\Delta$-MLFF data for all $\rho$-$T$ states: average ionization of the ions ($Z^*$), Coulomb coupling parameter ($\Gamma$), electron degeneracy parameter ($\Theta$), pressure ($P$), self-diffusion coefficient ($D$), shear viscosity ($\eta$), static ion structure factor ($S(\bm{q})$), and dynamic  ion structure factor ($S(\bm{q},\omega)$); comparison of the Mg pressure with the AIMD equation of state of Ref.~\onlinecite{gonzalez2020equation}.


\section*{Acknowledgements}
S.B. and P.S. gratefully acknowledge support from Grant No. DE-NA0004128 and DE-NA0004273 funded by the U.S. Department of Energy (DOE), National Nuclear Security Administration (NNSA). This work was performed, in part, under the auspices of the U.S. Department of Energy by Lawrence Livermore National Laboratory under Contract DE-AC52-07NA27344 and was supported by the LLNL-LDRD Program under Project No. 26-FS-026.

\section*{Author declarations}
The authors have no conflicts to disclose.

\section*{Data Availability Statement}
The data that support the findings of this study are available
within the article and its supplementary material.


\renewcommand{\bibsection}{\section*{References}}

%

\end{document}